\documentclass[prb,twocolumn,superscriptaddress,longbibliography]{revtex4-2}
\usepackage{amsmath,amssymb,bm}
\usepackage[
citecolor=ForestGreen,
colorlinks, 
linkcolor=violet,
urlcolor=black,]
{hyperref}
\usepackage{graphicx}
\usepackage{epstopdf}
\usepackage{latexsym}
\usepackage{subfigure}
\usepackage[usenames, dvipsnames]{color}
\usepackage[usenames, dvipsnames]{xcolor}
\usepackage{natbib}
\usepackage{braket}
\usepackage{float}
\usepackage[normalem]{ulem}
\usepackage{comment}
\usepackage{mathtools}
\usepackage{array}
\usepackage{tabu}
\usepackage{multirow}
\usepackage{chemformula}
\usepackage{svg}
\usepackage[T1]{fontenc}

\newcommand\redsout{\bgroup\markoverwith{\textcolor{red}{\rule[0.5ex]{2pt}{0.4pt}}}\ULon}
\newcommand\bluesout{\bgroup\markoverwith{\textcolor{blue}{\rule[0.5ex]{2pt}{0.4pt}}}\ULon}

\newcommand{\SPhide}[1]{{}}

\begin{document}

\title{A theorem on extensive ground state
entropy, spin liquidity and some related
models}
\author{Sumiran Pujari}
\affiliation{Department of Physics, Indian Institute of
Technology Bombay, Mumbai, MH 400076, India}
\affiliation{Max Planck Institute for the Physics of Complex Systems, 01187 Dresden, Germany}
\email{sumiran.pujari@iitb.ac.in}

\begin{abstract}
    An exact mechanism is written down to guarantee extensive residual ground state entropy and spin liquidity in spin-1/2 lattice models with bond-dependent couplings.
It is based on the presence of extensively large and mutually 
non-commuting (``\guillemotleft anticommuting\guillemotright'') sets of local conserved quantities with a gauge-like character. 
This mutual algebra is similar to those of spin-1/2 degrees of freedom however arising in the structure of \emph{local conserved charges} whose support is not restricted to a single lattice site.
The general theorem is first pedagogically illustrated through a variant of the familiar one-dimensional quantum Ising model featuring such an \guillemotleft anticommuting\guillemotright~structure.
This leads to classical spin liquidity co-existing with 
quantum Ising order.
The rest of the paper is then devoted to applications in higher dimensions with more general \guillemotleft anticommuting\guillemotright~structures which voids spin or magnetic ordering altogether.
Proofs of the resultant quantum spin liquidity are given through an analysis of static and dynamic $n$-point spin correlators relying solely on the \guillemotleft anticommuting\guillemotright~algebraic structure of the constructed models. 
It is not evident if they admit exact solutions using
known techniques. 
The precise nature of these quantum spin liquids is thus an open question including the existence of a quasiparticle description for these models. 
We compare and contrast them with other well-known quantum spin liquids.
\end{abstract}
\maketitle

\section{Introduction}
\label{sec:intro}

Exact statements are of immense value in quantum
many-body physics. They include exactly solvable
models of course, but also go beyond them. 
Well-known examples of the second kind are
the Peierls argument for classical Ising 
models~\cite{Peierls_1936,Bonati_2014} and
Elitzur's theorem in the context of lattice gauge 
theories that forbids local orders~\cite{Elitzur_1975}
with implications for the spontaneous symmetry breaking
in superconductors~\cite{Frohlich_Morchio_Strocchi_1981}. 
Other semi-rigorous to rigorous examples
are the Ginzburg criterion on the validity of mean-field
theories~\cite{Ginzburg_1961}, and Harris and Imry-Ma criteria 
on the effect of disorder on clean systems~\cite{Harris_1974,Vojta_Hoyos_2014,Imry_Ma_1975,Aizenman_Wehr_1989}. 

In this work,
we will make an exact statement of this second kind and
illustrate it through various models including
solvable ones. The statement concerns a 
theoretical mechanism that forces 
an extensive residual ground state 
entropy on a system along with quantum spin liquidity 
as the \emph{provable} physical consequences
as we shall see. Systems with extensive 
ground state entropy are 
often interesting with extremely correlated physics 
down to the lowest temperatures.
Well-known examples are classical spin ices~\cite{spin_ice_review} 
and the SYK model~\cite{SYK_review}. 
This may seem pathological and in violation
of the third law of thermodynamics~\cite{third_law}
to a novice in the field of strongly correlated matter. 
This is rather understood  to
be the physical state of affairs
for generic temperature scales~\cite{Masanes_Oppenheim_2017}
similar to classical spin ices
for example. 
In a ``realistic'' 
situation, other (even smaller) couplings will then select 
the “true” ground state 
to accord with 
the third law of thermodynamics often at inaccessible
temperatures from a practical point of view.
A well-known example of such an inaccessible physics
is the prediction of a crystalline state
by Wigner~\cite{Wigner_1934} in a 
jellium model of interacting electrons,
though there have been other physical
situations where this prediction has been
realized~\cite{Wigner_wiki}. Here we are
not going to concern ourselves with this 
issue, and are broadly going to focus on the
regime of extensive ground state entropy.

\subsection{Illustrative models in one dimension}
\label{sec:1d}

\begin{figure}
    \centering
    \includegraphics[width=0.6\linewidth]{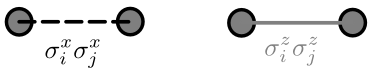} \\
    (a) \includegraphics[width=0.7\linewidth]{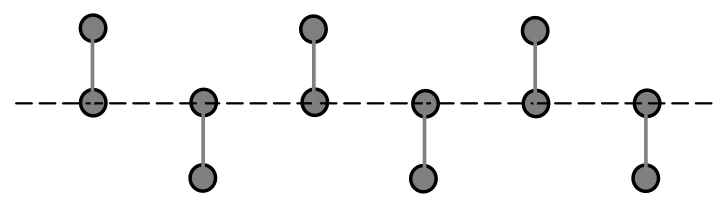}
    \vspace{3mm} \\
    (b) \includegraphics[width=0.7\linewidth]{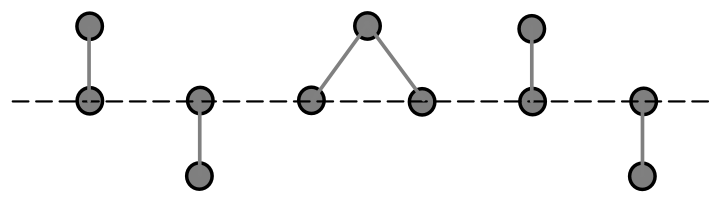} 
    \vspace{3mm} \\
    (c) \includegraphics[width=0.7\linewidth]{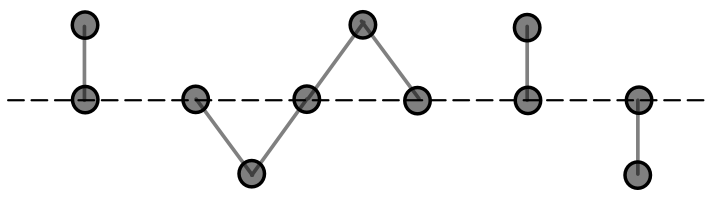}
    \vspace{3mm} \\
    (d) \includegraphics[width=0.7\linewidth]{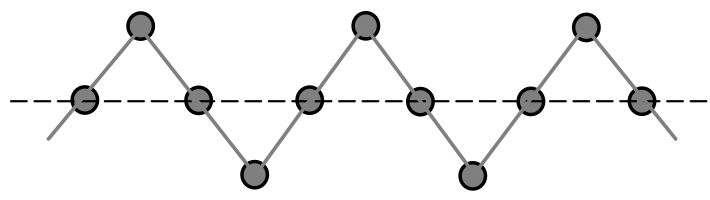} 
    \vspace{3mm} \\
    (e) \includegraphics[width=0.7\linewidth]{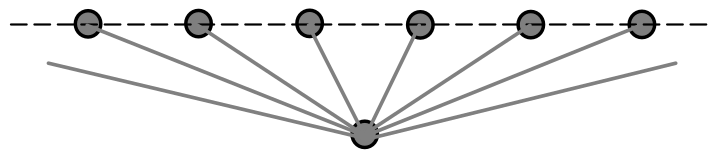}
    \caption{Examples of quantum Ising chains
    with different configurations for the
    auxiliary spins.}
    \label{fig:examples1}
\end{figure}

With the above motivation, we present
the basic ingredients of the theorem 
in a pedagogical fashion through a variant of
the familiar one-dimensional quantum Ising model. 
Consider the Hamiltonian
\begin{equation}
    H = J_x \sum_{\langle i,j \rangle} \sigma^x_i \sigma^x_j
+ J_z \sum_i \sigma^z_i \sigma^z_{\partial i}
\label{eq:H1}
\end{equation}
where $\partial i$ stands for the auxiliary partner
of site $i$ on the spin chain. 
It is as if we are ``applying'' the transverse field
--- of the standard transverse field quantum Ising model
(TFQIM) $J \sum_{\langle i,j \rangle} \sigma^x_i \sigma^x_j
+ h \sum_i \sigma^z_i$ ---
but now via a transverse Ising coupling of the spins to partner
auxiliary spins.
Several examples are
shown in Fig.~\ref{fig:examples1}. We will stick to
ferromagnetic couplings throughout in this article
without loss of generality.

Case (a) corresponds to when all sites on the chain
obey the unique auxiliary partner condition mentioned
in the abstract.
Case (b-e) corresponds to when not all sites on the chain
obey the unique auxiliary partner condition.
Case (d-e) corresponds to when all sites on the chain
violate the unique auxiliary partner condition.

In all cases, we have the standard global
$Z_2$ symmetry of the TFQIM. It may be implemented as
a $180^\circ$ rotation around the $z$-axis, i.e.
\begin{equation}
    \mathcal{U}^{Z_2} = \prod_i \otimes \; \mathcal{R}^{\pi,z}_i
\end{equation}
with
\begin{equation}
    \mathcal{R}^{\pi,z}_i = e^{i \pi \sigma^z_i/2}.
\end{equation}
Under this
\begin{align}
    \sigma^z_i & \rightarrow \mathcal{U}^{Z_2} \sigma^z_i {\mathcal{U}^{Z_2}}^\dagger = \mathcal{R}^{\pi,z}_i \sigma^z_i \mathcal{R}^{-\pi,z}_i = \sigma^z_i \\
    \sigma^x_i & \rightarrow - \sigma^x_i \\
    \text{and } H & \rightarrow H
\end{align}
The consequence of this is 
the conservation of the parity
of total chain magnetization in the $z$-direction 
$M^z = \sum_i \sigma^z_i$. Clearly, $[H,M^z] \neq 0$
but
\begin{align}
    [H,M^z\text{ mod }2] & = \sum_{\langle i,j \rangle} 
                    [\sigma^x_i \sigma^x_j, M^z \text{ mod }2] = 0
\end{align}  
This conservation is also equivalent to fermion
parity conservation after the Jordan-Wigner 
transformation~\cite{Jordan_Wigner_1928}
which maps the Ising term $\sigma^x_i \sigma^x_j$
to a sum of hopping and superconducting terms in
the fermion language.
Also in all cases, $\sigma^z_{\partial i}$ is conserved for
all $i$, i.e.
\begin{equation}
    [H,\sigma^z_{\partial i}] = 0
\end{equation}
as can be verified easily. Thus this degree of
freedom becomes effectively classical.
This in fact facilitates
the computation of the exact eigenspectrum via the
Jordan-Wigner transformation~\cite{JWpreprint,Nussinov_Ortiz_2008}.

Let us start with case (a) 
in Fig.~\ref{fig:examples1} which satisfies the unique
auxiliary partner condition for all sites of the chain. 
For this case, we have the following:
\begin{itemize}
    \item The degeneracy of the spectrum 
is $2^{N_{\partial i}}$ where $N_{\partial i}$
is the number of the auxiliary spins.
    \item Additionally, the eigenspectrum remains 
    the same as that of the quantum Ising model
with $\frac{h}{J} = \frac{J_z}{J_x}$. 
\end{itemize}

We can interpret the above as:
\begin{itemize}
    \item The auxiliary
spins remain paramagnetic down to zero temperature
co-existing with Ising order/disorder on the chain
~\cite{JWpreprint}.
    \item The presence of an extensive residual  
    entropy or finite residual
    entropy density.
    \item One may call this ground state as a
    co-existence state of Ising order/disorder
    with a ``classical'' spin liquid. We will prove
    this liquidity aspect in Sec.~\ref{subsec:liquidity}.
\end{itemize}

To prove the above we use the following lemma:
Let there be two conserved quantities
$A$ and $B$, i.e. $[H,A]=[H,B]=0$,
that are mutually anticommuting $\{A,B\}=0$.
For an eigenstate in the $A$-basis,
i.e. $H|\psi\rangle = E|\psi\rangle$ and 
$A|\psi\rangle = a|\psi\rangle$, there
exists another state $|B\psi\rangle
\equiv B|\psi\rangle$ which is also an eigenstate
with $H|B\psi\rangle = E|B\psi\rangle$ and 
$A|B\psi\rangle = -a|B\psi\rangle$. If $A$
has no zero eigenvalues, then $|B\psi\rangle$
is distinct than $|\psi\rangle$.

The proof goes as follows:
There are additional conserved quantities 
which are absent
in the TFQIM. These are $\sigma^x_i \sigma^x_{\partial i}$,
i.e.
\begin{equation}
    [\sigma^x_i \sigma^x_{\partial i} , H]=0
\end{equation}
for all $i$ as can be verified easily.
Furthermore these conserved quantities anticommute
with $\sigma^z_i$, i.e.
\begin{equation}
    \{\sigma^x_i \sigma^x_{\partial i}, \sigma^z_{\partial i}\}=0
\end{equation}
for all $i$ as can be verified easily.

As nomenclature, we will call sets of local 
conserved quantities with support over $O(1)$ sites which 
anticommute when they have site(s) in
common (and commute when no sites in common)
as ``\guillemotleft anticommuting\guillemotright''~sets
of local conserved quantitites.
We will also sometimes refer to them as an 
\guillemotleft anticommuting\guillemotright~structure, an \guillemotleft anticommuting\guillemotright~mechanism, or
an \guillemotleft anticommuting\guillemotright~algebra
of local conserved quantities. The above conserved sets
$\{\sigma^x_i \sigma^x_{\partial i}\}$ and $\{\sigma^z_{\partial i}\}$ form the first example of this structure in this
paper.

Also both sets of conserved quantities square to 
non-zero values and thus have no zero eigenvalues
\begin{align}
    \left(\sigma^z_i\right)^2 & = 1 \\
    \left(\sigma^x_i \sigma^x_{\partial i}\right)^2 & = 1
\end{align}
Thus by the application of the lemma above,
for each eigenstate $|\psi\rangle$ of $H$, one arrives
at $N_{\partial i}$ degenerate eigenstates
as $\left(\sigma^x_i \sigma^x_{\partial i} \right)|\psi\rangle$. In fact there are many
more degenerate eigenstates arrived at by the
operation of the product of $\left(\sigma^x_i \sigma^x_{\partial i} \right)$ over any subset
of the auxiliary partner sites. One can convince
oneself that the total degeneracy is 
thus $2^{N_{\partial i}}$.

The eigenspectrum is same as that of TFQIM
with $\frac{h}{J}=\frac{J_z}{J_x}$ can be
intuitively seen by choosing that sector
of the Hamiltonian which corresponds to
all the conserved $\sigma^z_{\partial i}$
being all up or all down, i.e.
$\prod_{\partial i} \otimes |\uparrow^z_{\partial i}\rangle$
or $\prod_{\partial i} \otimes |\downarrow^z_{\partial i}\rangle$. 
We formalize this as follows:
Let us look at a particular sector or block of
the Hamiltonian organized in the basis
of conserved quantities $\{\sigma^x_i \sigma^x_{\partial i} \}$, say corresponding to 
$\langle \sigma^x_i \sigma^x_{\partial i} \rangle = 1$
for all $i$.
For each given $\langle \sigma^x_i \sigma^x_{\partial i} \rangle$, there are two compatible states on the
bond $(i,\partial i)$. 
For $\langle \sigma^x_i \sigma^x_{\partial i} \rangle = 1$, we have the states $|\pm^x_i \pm^x_{\partial i}\rangle$ on $(i,\partial i)$ bond. 
For $\langle \sigma^x_i \sigma^x_{\partial i} \rangle = -1$, we have the states $|\pm^x_i \mp^x_{\partial i}\rangle$.
$\sigma^z_i \sigma^z_{\partial i}$ flips between
the two compatible states on $(i,\partial i)$ bond.
($\sigma^z_i \sigma^z_{\partial i} |\pm^x_i \pm^x_{\partial i} \rangle =
|\mp^x_i \mp^x_{\partial i} \rangle$ for
$\langle \sigma^x_i \sigma^x_{\partial i} \rangle = 1$,
and $\sigma^z_i \sigma^z_{\partial i} |\pm^x_i \mp^x_{\partial i}\rangle = |\mp^x_i \pm^x_{\partial i}\rangle$ for
$\langle \sigma^x_i \sigma^x_{\partial i} \rangle = -1$
respectively.) 
Thus $\sigma^z_i \sigma^z_{\partial i}$ terms  leads to off-diagonal
matrix elements for this form of Hamiltonian blocks. 
If the Hamiltonian blocks were to
be organized using the other conserved set $\{ \sigma^z_{\partial i}\}$, then $\sigma^z_i \sigma^z_{\partial i}$ would be a diagonal operator.
In our organization of the Hamiltonian blocks using
the conserved set $\{\sigma^x_i \sigma^x_{\partial i} \}$, 
the operator $\sigma^x_i \sigma^x_j$ on the nearest neighbour
bonds $(i,j)$ along the chain now measures the parity
of the $\sigma^x$-state on these bonds by definition and is
thus a diagonal operator.
($\sigma^x_i \sigma^x_j |\pm^x_i \pm^x_j \rangle =
+ |\pm^x_i \pm^x_j\rangle$
and $\sigma^x_i \sigma^x_j |\pm^x_i \mp^x_j \rangle =
- |\pm^x_i \mp^x_j\rangle$ respectively.) 
Thus Eq.~\ref{eq:H1} reduces to
an effective (dual) TFQIM once the value of 
$\langle \sigma^x_i \sigma^x_{\partial i} \rangle$ is
chosen on all $(i,\partial i)$ bonds. We may write
it as follows
\begin{equation}
    H = J^{\text{eff}} \sum_{\langle i,j \rangle} \tau^z_i \tau^z_j
+ h^{\text{eff}} \sum_i \tau^x_i 
\label{eq:dual_H1}
\end{equation}
where the $\tau$ operators operate on the two
states consistent with $\langle \sigma^x_i \sigma^x_{\partial i} \rangle$, and
$J^{\text{eff}} = J_x$, 
$h^{\text{eff}} = J_z$. As an aside, the 
parenthetical ``dual"
refers to the interchanging of the diagonal
and off-diagonal operations when organizing
the Hamiltonian blocks using the conserved
set $\{\sigma^z_{\partial i}\}$ 
versus the conserved set $\{\sigma^x_i \sigma^x_{\partial i}\}$, 
while choosing $z$-axis
to be the
spin-$\frac{1}{2}$ quantization axis
as is commonly done.

Now let us consider the case (b) in Fig.~\ref{fig:examples1}.
Here again we have the conservation of
$\sigma^x_i \sigma^x_{\partial i}$ for all $i$
with unique partners. For the two sites which share
a partner, the conserved quantity is now
$\sigma^x_i \sigma^x_{\partial (i,i+1)} \sigma^x_{i+1}$.
This also anticommutes with $\sigma^z_{\partial (i,i+1)}$.
Thus we can make similar arguments as above.
In case (c), 
the conserved quantity is 
$\sigma^x_i \sigma^x_{\partial (i,i+1)} \sigma^x_{i+1} 
\sigma^x_{\partial (i+1,i+2)} \sigma^x_{i+2}$ with
similar physics since in all the above cases (a-c)
there are an extensive number of additional conserved
quantities.
In case (d), the unique partner condition is lost
for the full spin chain. Thus in this case we 
do not have an extensive number of additional
conserved quantities. There is only one such quantity,
i.e.
$\prod_{i,\partial(i,i+1)} \otimes \sigma^x_i 
\otimes \sigma^x_{\partial(i,i+1)}$.
This will lead to a degeneracy of 2 of the
spectrum. The configuration of the auxiliary
spins which corresponds to the ground state
sector also needs determination~\cite{JWpreprint}.
Case (e) is another such example to show why
the physics present in cases (a-c) is absent
in the TFQIM. The operator in this case
is $ \sigma^x_{\partial} \otimes \prod_{i} \otimes \sigma^x_i $ 
which is reminiscent
of the string operator $\prod_{i} \otimes \sigma^z_i$
that measures the conserved parity of the 
magnetization in TFQIM.

Before proceeding further, 
we note that 
the above result in the context of quantum Ising models
is essentially a restatement of the result obtained 
in Ref.~\cite{Nussinov_Ortiz_2008} where the spectral
equivalence was rather shown in a converse fashion 
-- modulo minor details -- 
using the set of 
conserved quantities $\{\sigma^z_{\partial i}\}$ 
for the Hamiltonian block organization.
The context of Ref.~\cite{Nussinov_Ortiz_2008} was that
of quantum compass models with the motivation
coming from the physics of orbital degrees of freedom
in transition metal systems. 
The more general structure
of ``bond algebras'' underlying this work is further elaborated in Refs.~\cite{Nussinov_Ortiz_2009}, including resultant dualities in Refs.~\cite{Cobanera_Ortiz_Nussinov_2010,Cobanera_Ortiz_Nussinov_2011} an example of which we saw as Eq.~\ref{eq:dual_H1} in 
the analysis of Eq.~\ref{eq:H1} above.
Furthermore the lemma above and its variants have
been used previously in other contexts~\cite{Fendley_2016},
including the context of ground state degeneracy and 
topological orders. See in
particular Ref.~\cite{Nussinov_Ortiz_prb2023} which
gives a good overview of the existing results in the
literature.
The attention in these works has been on sub-extensive 
degeneracies and, generally speaking, topological (or
non-topological) quantum orders that have ground state
manifolds with zero ground state entropy density 
which is not the focus of this work.
The model Hamiltonians that
form the main subject of this paper
(Sec.~\ref{sec:2d})  are also 
Nussinov-Ortiz bond
algebras~\cite{Nussinov_Ortiz_2008,Nussinov_Ortiz_2009}.
However, the ingredient of the
\guillemotleft anticommuting\guillemotright~algebraic
structure of extensively many local conserved quantities 
leads to the new physics reported in this paper, i.e.
finite ground state entropy density and quantum
spin liquidity. We will compare and contrast
this with the physics of other known quantum spin liquids
in Sec.~\ref{subsec:compare_contrast}.

The various lattice geometries considered above in Fig.~\ref{fig:examples1} have been found in real material systems. 
In particular, the example of Fig.~\ref{fig:examples1}(a),(d) are referred to as the comb or branched chains and sawtooth chains respectively. 
See Ref.~\cite{Shvanskaya_Vasiliev_2024} and references therein for sawtooth chain examples.
For recent examples relating to comb lattices, see e.g. Ref.~\cite{Chepiga_White_2020} from a quantum criticality perspective, and Ref.~\cite{Bhattacharya_etal_2021} for a very different perspective concerning qubit regularization for the quantum simulation of quantum field theories~\cite{Chandrasekharan_plenary_2025}.
These examples though focus on $SU(2)$-symmetric situations generally speaking, or perhaps easy-plane or easy-axis $XXZ$ anisotropies but in a bond-independent way that are distinct from the bond-dependent coupling Hamiltonians that have been considered in this work. 
One may legitimately ask if an analog of the Jackeli-Khaliullin's mechanism~\cite{Jackeli_Khaliullin_PRL_2009} exists that would lead to the particular bond dependence considered in Eq.~\ref{eq:H1} or later in Sec.~\ref{sec:2d} from a quantum chemical perspective. 
This is not known to the author and is left to the future.

\subsubsection{Further discussion}
\label{subsec:1d_discussion}
Let us consider case (a) for this discussion.
It is natural to block diagonalize the Hamiltonian
$H$ in terms of the conserved spin configurations
of the auxiliary partner spins
$\prod_{\partial i} \otimes |\sigma^z_{\partial i} \rangle$.
However, 
the conservation of $\sigma^x_i \sigma^x_{\partial i}$ 
begs the following question: How to understand
the physics if we were to
organize the Hamiltonian blocks in terms
of the conserved $\sigma^x_i \sigma^x_{\partial i}$ 
for all $i$~?
Firstly, fixing the configuration of the auxiliary
spins  as 
$\prod_{\partial i} \otimes |\sigma^z_{\partial i} \rangle$
implies no fluctuation in them.
But fixing the eigenvalues (of $\pm 1$) of the
conserved $\sigma^x_i \sigma^x_{\partial i}$
for all $i$ does not imply any such thing.
In this way of block diagonalization, both
the spins of the spin chain and the auxiliary
spins keep fluctuating. This suggests that 
the (local) conservation of 
$\sigma^x_i \sigma^x_{\partial i}$
has a gauge-like character.
From this point of view, for a given eigenstate
$|\psi \rangle$, we can obtain degenerate eigenstates
as $\sigma^z_{\partial i} |\psi \rangle$
or $\prod_{\{\partial j\} \subseteq \{\partial i\}} \sigma^z_{\partial j} |\psi \rangle$ for any
subset of auxiliary spins. This again gives
a degeneracy of $2^{N_{\partial i}}$ as expected.
Due to this extensive degeneracy, the gauge
charges or eigenvalues of the conserved
$\sigma^x_i \sigma^x_{\partial i}$ can also
keep fluctuating. This is because we can linearly
combine the eigenstates from different gauge
charge sectors to obtain a new eigenstate. Under
time evolution, this linear combination will stay
put, i.e. both the gauge charges and the
auxiliary spin states keep fluctuating for all times.

Another perspective is to look at the same physics
after the Jordan-Wigner transformation. Then we arrive
at 
\begin{equation}
H = J_x \sum_{\langle i,j \rangle} 
\left(c^\dagger_i c_j + c^\dagger_i c^\dagger_j + \text{h.c. }\right)
+ J_z \sum_i (2 n_i - 1)(2 n_{\partial i} - 1)
\end{equation}
The global fermion parity is again conserved due
to global $Z_2$ symmetry,
but we also have local $Z_2$ symmetries
in terms of local $180^\circ$ rotations around
the $x$-axis for site $i$ and $\partial i$ which keep 
the Hamiltonian unchanged.
This implies the conservation of  
$\sigma^x_i \sigma^x_{\partial i}$
on $(i,\partial i)$ bonds.
Upon Jordan-Wigner transformation, we get
\begin{equation}
    [H, (-1)^\text{Jordan-Wigner phases} \left(c^\dagger_i c_{\partial i} + c^\dagger_i c^\dagger_{\partial i} + \text{h.c.}\right)]=0.
\end{equation}
However, there are no kinetic hopping or superconducting
terms $\propto \left(c^\dagger_i c_{\partial i} + c^\dagger_i c^\dagger_{\partial i} + \text{h.c.}\right)$ 
corresponding to the local $Z_2$ charges on
$(i,\partial i)$ bonds. Thus all gauge sectors are
degenerate. Also the mutual anticommutation of
\begin{equation}
    \{ c^\dagger_i c_{\partial i} + c^\dagger_i c^\dagger_{\partial i} + \text{h.c.},n_{\partial i} \} = 0
\end{equation}
implies that local $Z_2$ charges can fluctuate
along with $n_{\partial i}$.

\subsection{Some Extensions}
\label{subsec:1dextensions}
Let us continue with case (a). 
Since $\sigma^x_i \sigma^x_{\partial i}$ is conserved,
the following Hamiltonian 
\begin{equation}
    H = J_x \sum_{\langle i,j \rangle} \sigma^x_i \sigma^x_j
+ J_z \sum_i \sigma^z_i \sigma^z_{\partial i}
+ J'_x \sum_i \sigma^x_i \sigma^x_{\partial i}
\end{equation}
also is solvable. However $\sigma^z_{\partial i}$ is
not conserved anymore. Thus the extensive degeneracy
will be lost. The spectrum now will depend on the
conserved value of $\sigma^x_i \sigma^x_{\partial i}$ 
on all $(i,\partial i)$ bonds. E.g. the ground
state will correspond to 
$\langle \sigma^x_i \sigma^x_{\partial i} \rangle = 1$
for ferromagnetic $J'_x < 0$. 
Following same arguments on spectral equivalence
of Eq.~\ref{eq:H1} and the TFQIM as before in
Sec.~\ref{sec:1d} (see the discussion around Eq.~\ref{eq:dual_H1}), 
the above Hamiltonian also reduces to
an effective (dual) TFQIM once the value of 
$\langle \sigma^x_i \sigma^x_{\partial i} \rangle$ is
chosen on all $(i,\partial i)$ bonds. We may write
it as follows
\begin{equation}
    H = J^{\text{eff}} \sum_{\langle i,j \rangle} \tau^z_i \tau^z_j
+ h^{\text{eff}} \sum_i \tau^x_i 
+ J'_x \sum_i \langle \sigma^x_i \sigma^x_{\partial i} \rangle
\end{equation}
where the $\tau$ operators operate on the two
states consistent with $\langle \sigma^x_i \sigma^x_{\partial i} \rangle$ as befpre, and
$J^{\text{eff}} \propto J_x$, 
$h^{\text{eff}} \propto J_z$. One will again
obtain the TFQIM spectrum in any conserved sector.
The loss of extensive degeneracy corresponding
to $J'_x=0$ is seen through the third
term above $J'_x \sum_i \langle \sigma^x_i \sigma^x_{\partial i} \rangle$. One sees that 
there are still degenerate excited sectors given
by different configurations of $\langle \sigma^x_i \sigma^x_{\partial i} \rangle$
which keep the sum $\sum_i \langle \sigma^x_i \sigma^x_{\partial i} \rangle$ fixed. The degeneracies
are basically $N_{i} \text{ choose } N_{\langle \sigma^x_i \sigma^x_{\partial i} \rangle=1}$.
These degeneracies will likely be further broken down
in presence of additional solvability breaking terms. 
By a similar token, 
for the following Hamiltonian
\begin{equation}
    H = J_x \sum_{\langle i,j \rangle} \sigma^x_i \sigma^x_j
    + J'_z \sum_{\langle i,j \rangle} \sigma^z_i \sigma^z_j
+ J_z \sum_i \sigma^z_i \sigma^z_{\partial i}
\end{equation}
$\sigma^x_i \sigma^x_{\partial i}$ is not
conserved anymore,
but $\sigma^z_{\partial i}$ stays conserved. Thus
the extensive degeneracy will again be lost. 
\begin{figure}
    \centering
    \includegraphics[width=0.65\linewidth]{couplings.png} \\
    \includegraphics[width=0.9\linewidth]{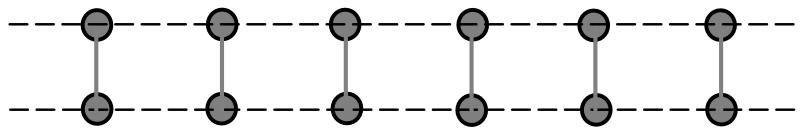}
    \caption{Ladder geometry with bond-dependent couplings
as discussed in the text.}
    \label{fig:example2}
\end{figure}
However,
solving for the spectrum using the Jordan-Wigner
transformation is not that straightforward. 

Also
may be noted that the following related
ladder Hamiltonian
\begin{align}
    H = & \;\; J_x \sum_{\langle i,j \rangle} \sigma^x_i \sigma^x_j
    + J_{\partial x} \sum_{\langle i,j \rangle} \sigma^x_{\partial i} \sigma^x_{\partial j} 
    \nonumber \\
 & + J_z \sum_i \sigma^z_i \sigma^z_{\partial i}
+ J'_x \sum_i \sigma^x_i \sigma^x_{\partial i}
\end{align}
as shown in Fig.~\ref{fig:example2} is effectively 
equivalent to
\begin{equation}
    H = \sum_{\langle i,j \rangle} J^{\text{eff}}_{ij}  \tau^z_i \tau^z_j
+ h^{\text{eff}} \sum_i \tau^x_i 
+ J'_x \sum_i \langle \sigma^x_i \sigma^x_{\partial i} \rangle
\end{equation}
with $h^{\text{eff}} \propto J_z$.
The case of $J^{\text{eff}}_{ij}$ requires more attention.
For the (ground state) sector corresponding to
$\langle \sigma^x_i \sigma^x_{\partial i} \rangle=1$ on
all $(i,\partial i)$ bonds, $J^{\text{eff}}_{ij} \propto (J_x + J_{\partial x})$
independent of the bond location.
The same would be true for the sector corresponding
to $\langle \sigma^x_i \sigma^x_{\partial i} \rangle=-1$ on
all $(i,\partial i)$ bonds. Recall we are considering ferromagnetic
couplings in this article throughout.
For other sectors where $\langle \sigma^x_i \sigma^x_{\partial i} \rangle$
is not uniformly the same sign, the bond location becomes important.
For a bond $(i,j)$ such that $\langle \sigma^x_i \sigma^x_{\partial i} \rangle=
\langle \sigma^x_j \sigma^x_{\partial j} \rangle$,
$J^{\text{eff}}_{ij} \propto (J_x + J_{\partial x})$.
For a bond $(i,j)$ such that $\langle \sigma^x_i \sigma^x_{\partial i} \rangle
\neq
\langle \sigma^x_j \sigma^x_{\partial j} \rangle$,
$J^{\text{eff}}_{ij}$ itself fluctuates between $\propto (J_x - J_{\partial x})$
and $\propto -(J_x - J_{\partial x})$ depending on the
state of the spins on the $(i,\partial i)$ and $(j,\partial j)$ 
bonds. Obtaining the spectrum in these excited sectors is
therefore more involved. For $J_x = J_{\partial x}$~\cite{Brzezicki_Oles_2009} which
would be the case in presence of mirror symmetry between
the two legs of the ladder, there
occurs a simplification and $J^{\text{eff}}_{ij} = 0$ on those
bonds where $\langle \sigma^x_i \sigma^x_{\partial i} \rangle
\neq
\langle \sigma^x_j \sigma^x_{\partial j} \rangle$. This leads
to disconnected TFQIM segments which can again be solved for
the excited eigenspectrum. The resultant  
physics has been more generally termed as Hilbert space 
fragmentation~\cite{Patil_Sandvik_2020}.
The above ladder Hamiltonian can be looked
at as a solvable local spinless fermionic model
for $J'_x=0$,
\begin{align}
    H  = & \;\; -t_x \sum_{\langle i,j \rangle} 
    (c^\dagger_i c_j + c^\dagger_i c^\dagger_j + \text{ h.c.}) 
        \\
        & - t_{\partial x} \sum_{\langle i,j \rangle} 
    (c^\dagger_{\partial i} c_{\partial j} + c^\dagger_{\partial i} c^\dagger_{\partial j} + 
    \text{ h.c.}) 
 + V \sum_i n_i n_{\partial i}  \nonumber
\end{align}
The physical situation is that of two 
$p$-wave superconducting wires coupled through
a short-ranged Couloumb interaction. Analogous
physics will carry through in this context.
We may conjecture that the physics extends to 
situation when the hopping and pairing amplitudes
are not exactly equal. A stronger conjecture would
be the stability of the ground state in presence
of hopping and/or superconducting amplitudes
between the two wires. The ground state of the 
fermionic system will be two locked superconducting
ground states independent of $V$.
The solvable case of $J'_x \neq 0$ leads to
non-local terms in the fermionic situation and may
be ignored. 
We end this section with the note that the quasi-one-dimensional ladder geometry considered in this section is perhaps more well-known than the comb or sawtooth geometries seen previously. 
Lattice models with ladder geometry are quite relevant for experiments and a great variety of them have been studied in both spin and fermionic contexts with various perspectives and motivations~\cite{1d_magnetism_review_2004, 1d_magnetism_review_2018}.

\section{Constructions in two dimensions}
\label{sec:2d}

Till now the discussion has been limited to
one dimensional systems. Let us now construct 
two-dimensional spin models with extensive ground
state entropy guaranteed through the mechanism
underlying the theorem, i.e. existence of 
extensively large mutually
\guillemotleft anticommuting\guillemotright~sets of  
conserved quantities. 
To repeat the nomenclature introduced in 
Sec.~\ref{sec:1d}, by
\guillemotleft anticommuting\guillemotright~sets, we mean 
the anticommutation of two conserved operators coming 
from different sets when they have a common site between 
them as in the previous Sec.~\ref{sec:1d} and throughout
the paper. The condition of a single common site is however
not strict, even though respected in all the models discussed
in this paper. One can easily construct variants 
where one may have the \guillemotleft anticommuting\guillemotright~mechanism operational 
even when local conserved quantities
coming from different sets have more than one site in common. Conserved quantities without
common sites of course keep commuting in the context
of spin models. We continue to take
ferromagnetic signs for the bond-dependent couplings.

\subsection{Generic degeneracy counting}
\label{subsec:counting}

\begin{figure}
    \centering
    \includegraphics[width=0.65\linewidth]{couplings.png} \\
    \includegraphics[width=0.7\linewidth]{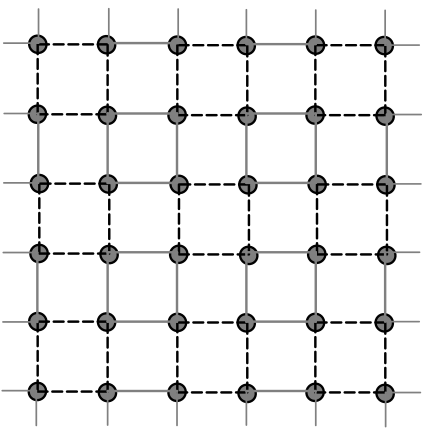} 
    \includegraphics[width=0.7\linewidth]{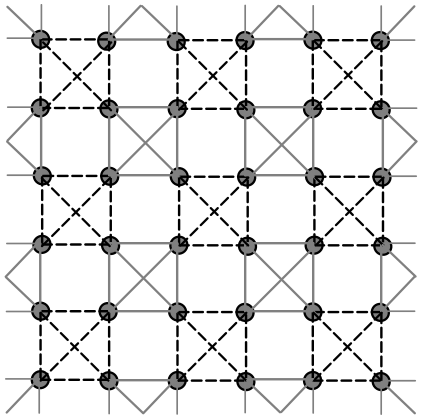}
    \caption{Prototypical two-dimensional spin models with
    extensive ground state entropies.}
    \label{fig:2dmodela}
\end{figure}

Consider the bond-dependent Hamiltonian 
\begin{equation}
    H =\;  \sum_{\boxed{x}} \left( \sum_{\langle i,j \rangle
    \in\; \boxed{x}} J^x_{ij} \sigma^x_i \sigma^x_j \right)
    +
    \sum_{\boxed{z}} \left( \sum_{\langle i,j \rangle
    \in\; \boxed{z}} J^z_{ij} \sigma^z_i \sigma^z_j \right).
    \label{eq:2dmodela}
\end{equation}
The bipartite nature of the lattice is not a requirement, and one may also consider a non-bipartite version of the above Hamiltonian in Eq.~\ref{eq:2dmodela} with next-nearest neighbour couplings included, i.e.,
\begin{equation}
    H =\;  \sum_{\boxed{x}} \left( \sum_{i,j
    \in\; \boxed{x}} J^x_{ij} \sigma^x_i \sigma^x_j \right)
    +
    \sum_{\boxed{z}} \left( \sum_{i,j
    \in\; \boxed{z}} J^z_{ij} \sigma^z_i \sigma^z_j \right).
    \label{eq:2dmodela_nonbipartite}
\end{equation}
Translationally invariant forms of the above Hamiltonians are sketched in Fig.~\ref{fig:2dmodela}. 
The systems are composed of square plaquettes with either $\sigma^x_i \sigma^x_j$ or $\sigma^z_i \sigma^z_j$ couplings exclusively arranged in a checkerboard like pattern. 
$\boxed{x}$ and $\boxed{z}$ denote the plaquettes with $\sigma^x_i \sigma^x_j$ and $\sigma^z_i \sigma^z_j$ respectively. 
Let us look at the conserved quantities. 
They are
\begin{enumerate}
    \item $\sigma^z_i \sigma^z_j \sigma^z_k \sigma^z_l$ 
    on the $\boxed{x}$ plaquettes.
    \item $\sigma^x_i \sigma^x_j \sigma^x_k \sigma^x_l$ 
on the $\boxed{z}$ plaquettes.
\end{enumerate}
The conserved nature of these quantities
may be verified easily. All the above quantities 
form extensively 
large sets due to their local nature. 
The two sets \guillemotleft anticommute\guillemotright~with 
each other. 

Eigenspectrum solvability of this model
is not apparent, but by the application of the lemma,
we can conclude that this system will host 
an extensive ground state entropy. In fact, the full
eigenspectrum will be massively degenerate in this
sense. The counting
can be ascertained by first spanning the system
with one of the conserved sets from the 
above that can serve as the basis for 
block-diagonalization of the Hamiltonian, and then
counting the other set that 
\guillemotleft anticommutes\guillemotright~with 
the chosen set. Thus
the ground state entropy is $\ln 2$ per unit cell.
A related Hamiltonian with the same degeneracy counting could
be constructed using the above conserved operators
directly,
\begin{equation}
    H =\; J_x \sum_{\boxed{x}} \left( \prod_{i\in\; \boxed{x}} \otimes \sigma^x_i \right) +
    J_z \sum_{\boxed{z}} \left( \sum_{i
    \in\; \boxed{z}} \otimes \sigma^z_i \right)
    \label{eq:2dmodelaprime}
\end{equation}
This would be in the spirit of Kitaev toric code~\cite{Kitaev_2003}
whose Hamiltonian is also the sum of the conserved
quantities which however mutually commute. We will
stick to models with two-spin terms without any
loss of generality.

\begin{figure}
    \centering
    \includegraphics[width=0.65\linewidth]{couplings.png} \\
    \includegraphics[width=0.7\linewidth]{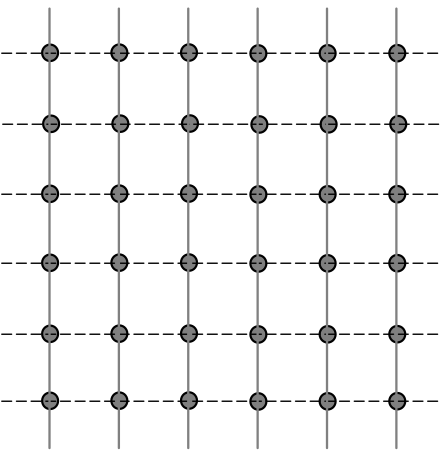}
    \caption{The $90^\circ$ compass model with
    only a double degeneracy via the anticommuting
    mechanism. The low-energy manifold is only
    sub-extensive in size, which
    leads to a zero low-energy entropy density
    in contrast to the other cases studied in this
    paper.}
    \label{fig:2dmodelb}
\end{figure}

An alternative Hamiltonian with this anticommuting
mechanism in operation is shown in Fig.~\ref{fig:2dmodelb}.
The system is now composed of crisscrossing Ising
chains with couplings in perpendicular directions
in spin space. It is a square lattice variant
of the Kitaev honeycomb model~\cite{Kitaev_2006} 
and in fact 
belongs to the class of ``compass'' 
models~\cite{Kugel_Khomskii_1973,Kugel_Khomskii_1982}.
Its ground state properties have been 
discussed in the literature~\cite{Doucot_etal_2005,Dorier_Becca_Mila_2005,Nussinov_vandenBrink_review_2015}. 
It can be written as
\begin{equation}
    H = \sum_{\mathbf{r}} J_x \sigma^x_{\mathbf{r}} \sigma^x_{\mathbf{r}+\mathbf{e_x}} + 
    J_z \sigma^z_{\mathbf{r}} \sigma^z_{\mathbf{r}+\mathbf{e_z}}
    \label{eq:2dmodelb}
\end{equation}

The degeneracy
counting in this model has also been done through
the lens of the anticommuting mechanism, however
the it only leads to a double degeneracy
independent of system
size~\cite{Doucot_etal_2005}. This is because the
conserved $Z_2$ parities for each Ising chain 
are \emph{non-local} string operators as in TFQIM:
$\prod_{r_y} \sigma^x_{(r_x,r_y)}$ for a given
$r_x$ and $\prod_{r_x} \sigma^z_{(r_x,r_y)}$ for a given
$r_y$. Note the number of these non-local 
conserved quantities is sub-extensive and not
extensive in contrast to the other models.
Due to the geometry of the
string operators -- each string operator in a
given direction intersects all string operators
in the perpendicular direction -- the application
of the lemma only gives a double degeneracy.
An interesting
counterpoint in the context of this paper is the following:
even though the degeneracy is $O(1)$ by
the anticommuting mechanism, it has been stated
by Dorier \textit{et al} that, ``When $J_x \neq J_z$, we show that, on clusters of dimension $L \times L$, the low-energy spectrum consists of $2^L$ states which collapse onto each other exponentially fast with $L$, a conclusion that remains true arbitrarily close to $J_x = J_z$. At that point, we show that an even larger number of states collapse exponentially fast with $L$ onto the ground state, and we present numerical evidence that this number is precisely $2\times2^L$.''~\cite{Dorier_Becca_Mila_2005}
It is as if the system ``would prefer'' a 
(sub-extensively) large
degeneracy, however there are no symmetries
to guarantee it exactly. In all the other constructions
discussed in this paper, we can guarantee rather
an extensive degeneracy because of the local nature
of the conserved quantities, i.e. they have support
on $O(1)$ lattice sites. Furthermore,
additional
degeneracies at ``fine-tuned'' coupling values
might be present even in these models, but we
are not concerning ourselves with such effects in
this paper. The generic extensive entropy
case 
afforded by the anticommuting mechanism of this 
paper can already be used to prove spin liquid
nature of these models
as will be discussed in the next Sec.~\ref{subsec:liquidity}.
Another example is a one-dimensional version of the 
compass model~\cite{Brzezicki_Dziarmaga_Oles_2007} 
where, by virtue of the reduced 
dimensionality, the conserved
quantities become local and the theorem then 
guarantees an extensive degeneracy even for 
finite chains~\cite{footnote_BDO07}.
It can also be reduced to an effective
TFQIM and follows the spirit of the one-dimensional 
cases discussed earlier.

\begin{figure}
    \centering
    \includegraphics[width=0.65\linewidth]{couplings.png} \\
    \includegraphics[width=0.8\linewidth]{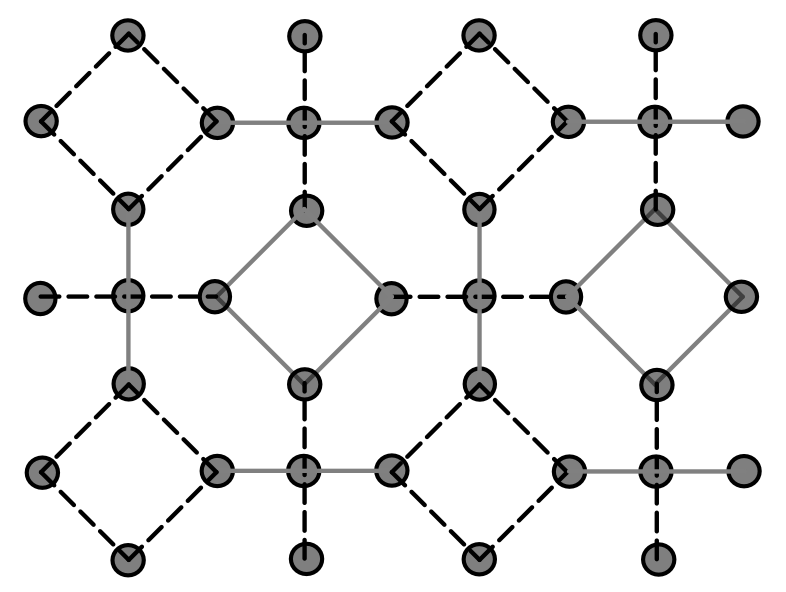}
    \caption{A more intricate two-dimensional 
    spin model with
    extensive ground state entropy.}
    \label{fig:2dmodel}
\end{figure}

Consider finally the system sketched in Fig.~\ref{fig:2dmodel} which has a more intricate structure compared to the previous models. 
This case is of interest because of the technical differences in its ground state entropy counting compared to before which might be worth pointing out.
As with Eq.~\ref{eq:2dmodela_nonbipartite} considered earlier, one may include next-nearest neighbour couplings on the square plaquettes of Fig.~\ref{fig:2dmodel} without lost of generality.
The system is now composed of a) square plaquettes with either $\sigma^x_i \sigma^x_j$ or $\sigma^z_i \sigma^z_j$ couplings exclusively denoted as $\boxed{x}$ and $\boxed{z}$ respectively, b) crosses ($\times$) composed of both $\sigma^x_i \sigma^x_j$ and $\sigma^z_i \sigma^z_j$ segments crisscrossing each other, and c) hexagonal plaquettes with alternating $\sigma^x_i \sigma^x_j$ and $\sigma^z_i \sigma^z_j$ couplings as a result of a) and b).
It can be formally written as
\begin{align}
    H = & \;\; \sum_{\boxed{x}} \sum_{\langle i,j \rangle
    \in\; \boxed{x}} J^x_{ij} \sigma^x_i \sigma^x_j + 
     \sum_{\boxed{z}} \sum_{\langle i,j \rangle
    \in\; \boxed{z}} J^z_{ij} \sigma^z_i \sigma^z_j 
    + \nonumber \\
    & J'_x \sum_{\times} \sum_{\langle i,j \rangle_x
    \in\; \times} \sigma^x_i \sigma^x_j
    + J'_z \sum_{\times} \sum_{\langle i,j \rangle_z
    \in\; \times} \sigma^z_i \sigma^z_j
    \label{eq:2dmodel}
\end{align}
for the variant with only nearest-neighbour couplings.
The conserved quantities are 
\begin{enumerate}
    \item $\sigma^z_i \sigma^z_j \sigma^z_k \sigma^z_l$ 
    on the $\boxed{x}$-plaquettes.
    \item $\sigma^x_i \sigma^x_j \sigma^x_k \sigma^x_l$ 
on the $\boxed{z}$-plaquettes.
    \item $\sigma^z_i \sigma^z_j \sigma^z_k$ 
on the bonds $\langle i,j \rangle_x \in\; \times$.
    \item $\sigma^x_i \sigma^x_j \sigma^x_k$ 
on the bonds $\langle i,j \rangle_z \in\; \times$.
\end{enumerate}
The conserved nature of these quantities
may be verified easily. There does not seem to be
any obvious conserved quantity associated with the 
hexagonal
plaquettes. All the above quantities form extensively 
large sets. The first and second sets of conserved 
quantities commute with each other. The third and
fourths sets \guillemotleft anticommute\guillemotright~with 
each other. 
Similarly, the first and fourth sets anticommute 
with each other and the second and third sets
\guillemotleft anticommute\guillemotright~with each other.
The first and third sets commute with each other,
and the second and fourth sets commute with each
other. 

Similar to Eq.~\ref{eq:2dmodela}, eigenspectrum solvability 
of the model in Eq.~\ref{eq:2dmodel}, Fig.~\ref{fig:2dmodel}
is not apparent and there will be a
massive degeneracy of its eigenspectrum. The counting
can be ascertained by first spanning the system
with mutually conserved sets from the above, and then
counting the remaining sets that 
\guillemotleft anticommute\guillemotright~with the spanning sets. The 
maximum of all possible ways of doing this will
give the entropy due to this mechanism. In this
model, if we use the first
and third sets as the spanning sets,
then the remaining sets contribute an entropy
of $3 \ln 2 \; k_B$ per unit cell.
Similarly, if we use second
and fourth sets as the spanning sets,
then the remaining sets again contribute an entropy
of $3 \ln 2 \; k_B$ per unit cell.
Instead, if we use the first and second sets 
as the spanning 
sets, then the remaining sets contribute an entropy of
$4 \ln 2 \; k_B$ per unit cell. 
The ground state entropy is therefore 
$4 \ln 2 \; k_B$ per unit cell through this 
anticommuting mechanism.

There is an unresolved puzzle with respect to
the above degeneracy counting even though it is
exact and non-perturbative. If we were to think
of Eq.~\ref{eq:2dmodel} through a perturbative
lens, then there are two natural ways of going about it.
If we take all the terms with $\sigma^x_i \sigma^x_j$
Ising couplings as the dominant terms and the
terms with $\sigma^z_i \sigma^z_j$
Ising couplings as the perturbation or vice versa,
one arrives at an entropy of 3 $\ln 2$ per unit cell.
On the other hand, if we take all the terms involving
the boxed plaquettes $\boxed{x}$ and $\boxed{z}$ as 
the dominant terms with the rest being the 
perturbation, then we arrive at an entropy of
4 $\ln 2$ per unit cell. Since 4 $\ln 2$ is the 
non-perturbative count, the additional $\ln 2$ 
contribution is not accounted for when setting up
the perturbation theory in the first manner. It is
a puzzle as to how this additional degeneracy would
be accounted for at all orders in perturbation 
theory when
doing it in this manner. Note that additional
degeneracies can arises at special points as pointed
out by Dorier \emph{et al}~\cite{Dorier_Becca_Mila_2005},
however here it must happen without any such
fine-tuning, i.e. for any value of the perturbation.

On a related note, for the model of Eq.~\ref{eq:2dmodela},
in the limit of $J_x=0$ or $J_z=0$ the degeneracy
corresponding to $\ln 2$ entropy per unit cell is
obvious. How this degeneracy survives to all orders
in perturbation theory when going away from these
limits is another related question. The
answer must be of the form that no perturbation at
any order can connect two degenerate states to
split the degeneracy. For an example of such an
effect, see e.g. Ref.~\cite{Schafer_etal_2023}.
This is a different question from the earlier puzzle.
In the earlier puzzle, we are asking how \emph{extra} 
degeneracies are \emph{generated} at all orders
in perturbation theory set up in a particular way. 
In the above question, we are asking how the 
\emph{already present} degeneracies are 
\emph{preserved} to all orders in perturbation theory.

\subsection{Quantum spin liquidity}
\label{subsec:liquidity}

We will now prove ground state spin liquidity
in the extensively degenerate models using 
just the anticommuation structure.
For example for the model of Fig.~\ref{fig:2dmodel}
or Eq.~\ref{eq:2dmodel},
there are three kinds of sites: 
sites at the centre of the crosses ($i_\times$), 
those on the $\boxed{x}$-plaquettes ($i_x$) 
and those on the $\boxed{x}$-plaquettes ($i_z$). 
The proof can be understood by
taking one representative example,
say the ground state expectation 
$\langle \sigma^\mu_{i_\times} \sigma^\nu_{j_\times}\rangle$
on two different faraway sites. This ground state expectation
value is to be understood as a thermal mixture over
the ground state manifold as $T \rightarrow 0$,
i.e. 
\begin{equation}
    \langle O \rangle(T \rightarrow 0) =
    \sum_{|\psi\rangle \in \{|\psi_{\text{gs}}\rangle \}}
    \langle \psi | O | \psi \rangle
    \label{eq:gsexpectation}
\end{equation}
where $\{|\psi_{\text{gs}}\rangle \}$ is the ground state
manifold.

Working in the basis of the first and third
commuting sets (``$z$''-basis), we can 
subdivide the ground
state manifold into distinct sets or 
``equivalence classes'' containing 16 ground states each
given the two unit cells to which the sites 
$i_\times$ and $j_\times$ belong.
If a generic set $A$ is indexed by a representative
ground state $|\psi^A_{\text{gs}}\rangle$, then we 
can generate
the other 15 ground states by the application of 
$\sigma^x_i \sigma^x_j \sigma^x_k \sigma^x_l$ and the two
different $\sigma^x_i \sigma^x_j \sigma^x_k$ belonging to
the two unit cells on $|gs_A\rangle$. 
(16=1+3+3+(3$\times$3).) If we sum 
$\langle \sigma^\mu_{i_\times} \sigma^\nu_{j_\times}\rangle$ over
all these sixteen states, one finds that the sum is
zero for all cases of $\mu,\nu$ except for
$\langle \sigma^x_{i_\times} \sigma^x_{j_\times} \rangle$. To show
that the sum is zero even in this case, one can rework
the above starting from ``$x$''-basis involving the second and
fourth sets.
Thus this will be true for
the overall ground state manifold sum. 

\begin{table*}[t]
 \caption{A generic set $A$ of 4 ground states that form
 an equivalence class given two sites $i_\times$ and
 $j_\times$ for the model of Eq.~\ref{eq:2dmodel} and
 Fig.~\ref{fig:2dmodel}.
 \label{tab:1}}
\begin{tabular}{||c | c | c | c||} 
 \hline
 $|\psi\rangle$ & $\langle \psi | \sigma^z_{\partial_1 i_\times} \sigma^z_{i_\times}
\sigma^z_{\partial_2 i_\times} | \psi \rangle$ & 
$\langle \psi | \sigma^z_{\partial_1 j_\times} \sigma^z_{j_\times}
\sigma^z_{\partial_2 j_\times} | \psi \rangle$ & 
$\langle \psi | \sigma^\mu_{i_\times} \sigma^\nu_{j_\times} | \psi \rangle$ \\ [0.5ex] 
 \hline\hline
 $|\psi^A_{\text{gs}}\rangle$ & +1 & +1 & $\langle \psi^A_{\text{gs}} |  \sigma^\mu_{i_\times} \sigma^\nu_{j_\times} | \psi^A_{\text{gs}} \rangle$\\ 
 \hline
$\sigma^x_{\partial_3 i_\times} \sigma^x_{i_\times}
\sigma^x_{\partial_4 i_\times} |\psi^A_{\text{gs}}\rangle$ & -1 & +1 & $\left(2 \delta_{\mu x} - 1\right) \langle  \psi^A_{\text{gs}} |  \sigma^\mu_{i_\times} \sigma^\nu_{j_\times} | \psi^A_{\text{gs}} \rangle$ \\ 
 \hline
 $\sigma^x_{\partial_3 j_\times} \sigma^x_{j_\times}
\sigma^x_{\partial_4 j_\times} |\psi^A_{\text{gs}}\rangle$ & +1 & -1 & $ \left(2\delta_{\nu x} - 1\right) \langle  \psi^A_{\text{gs}} |  \sigma^\mu_{i_\times} \sigma^\nu_{j_\times} | \psi^A_{\text{gs}} \rangle$ \\
 \hline
 $\sigma^x_{\partial_3 i_\times} \sigma^x_{i_\times}
\sigma^x_{\partial_4 i_\times} 
\sigma^x_{\partial_3 j_\times} \sigma^x_{j_\times}
\sigma^x_{\partial_4 j_\times} |\psi^A_{\text{gs}}\rangle$ & -1 & -1 & $\left(2\delta_{\mu x} -1\right)
\left(2\delta_{\nu x} - 1\right) \langle  \psi^A_{\text{gs}} |  \sigma^\mu_{i_\times} \sigma^\nu_{j_\times} | \psi^A_{\text{gs}} \rangle$ \\
 \hline
\end{tabular}
\end{table*}

We will present a simpler argument below
by only involving the conserved 
operators that include the sites $i_\times$ and
$j_\times$ which would lead to a
set of 4 ground states. The division into the 
set of 16
related ground states organized by
unit cells is somewhat more natural.
Working in the ``$z$''-basis,
let the representative state $|\psi^A_{\text{gs}}\rangle$ 
correspond to the value of +1 for the conserved quantities
$\sigma^z_{\partial_1 i_\times} \sigma^z_{i_\times}
\sigma^z_{\partial_2 i_\times}$
and
$\sigma^z_{\partial_1 j_\times} \sigma^z_{j_\times}
\sigma^z_{\partial_2 j_\times}$
connected to the two sites $i_\times$ and $j_\times$.
The set $A$ can be indexed by the values of all the
other conserved quantities in the ``$z$''-basis
\emph{given} the previous statement.
We arrive at Table~\ref{tab:1} after generating the
set of 4 states. The table shows that this
grouping of the ground state manifold is unique. This is
because the three states generated from $|\psi^A_{\text{gs}}\rangle$ can not correspond 
by construction to a different 
representative state since the
value of $\sigma^z_{\partial_1 i_\times} \sigma^z_{i_\times}
\sigma^z_{\partial_2 i_\times}$
and
$\sigma^z_{\partial_1 j_\times} \sigma^z_{j_\times}
\sigma^z_{\partial_2 j_\times}$
are not +1 for these three states.
Clearly the sum over these 4 states
is zero whenever $\mu \neq x$ or $\nu \neq x$. For
$\langle  \sigma^x_{i_\times} \sigma^x_{j_\times} \rangle$, we start in the ``$x$''-basis and redo the
above as mentioned before. The associated table would
be the same as Table~\ref{tab:1} with the interchanging
of $x$ and $z$ everywhere.

One can similarly argue for the vanishing of ``2-point'' spin order when $i_x$ or $i_z$ type of sites are involved.
Also, one can see from these arguments that the ``faraway'' requirement of the two sites is not very strict.
This is analogous to the Kitaev model~\cite{Baskaran_Mandal_Shankar_PRL_2007,
Nussinov_Chen_2008} however guaranteed through the anticommuting mechanism. 
In fact, as we see, we did not need any other representation (fermionic or otherwise) to prove this which highlights the reach of the anticommuting structure.
Furthermore, one can extend these arguments to ``$n$-point'' spin orders involving different unit cells. 
A similar argument goes through for the model in Eq.~\ref{eq:2dmodela} or Eq.~\ref{eq:2dmodela_nonbipartite} of Fig.~\ref{fig:2dmodela} which is composed of only
one kind of lattice site.

Multi-spin order parameter
correlations such as bond energies, plaquette spin products, etc. can 
survive the above cancellations. 
Let us restrict ourselves to the situation of
faraway unit cells and the model of
Eq.~\ref{eq:2dmodela} for simplicity. 
We will state the result without giving 
the proof. The proof logic follows from the 
above sort of arguments.
For a 2-spin operator 
$\sigma^\mu_i \sigma^\nu_j$ on a bond 
$\langle i,j \rangle$, its 2-bond correlators 
$\langle \left( \sigma^\mu_i \sigma^\nu_j \right) \left( \sigma^\gamma_k \sigma^\delta_l \right)\rangle$
are non-zero only when $\{\mu,\nu,\gamma,\delta\}$ 
correspond to the spin space indices
that \emph{appear} in the Hamiltonian on their
corresponding sites $\{i,j,k,l\}$. E.g., for 
bonds $\langle i,j \rangle$ and $\langle k,l \rangle$
which host $\sigma^x_i \sigma^x_j$ and 
$\sigma^x_k \sigma^x_l$, the only non-zero correlator
is $\langle \left( \sigma^x_i \sigma^x_j \right) \left( \sigma^x_k \sigma^x_l \right)\rangle$ while all other
correlators are zero through the above kind of
cancellation arguments.
Similarly for 
bonds $\langle i,j \rangle$ and $\langle k,l \rangle$
which host $\sigma^x_i \sigma^x_j$ and 
$\sigma^z_k \sigma^z_l$, the only non-zero correlator
is $\langle \left( \sigma^x_i \sigma^x_j \right) \left( \sigma^z_k \sigma^z_l \right)\rangle$ and so on.

For a 3-spin and higher-spin operators, one has
to pay a little more care. The 3-spin case exposes
the general structure of these multi-spin correlations.
For $\sigma^\mu_i \sigma^\nu_j \sigma^\gamma_k$ 
on 3 contiguous sites 
$\langle i,j,k \rangle$, its 2-bond correlators 
$\langle \left( \sigma^\mu_i \sigma^\nu_j \sigma^\gamma_k \right) \left( \sigma^\delta_l \sigma^\alpha_m \sigma^\beta_n \right)\rangle$
are non-zero only when $\{\mu,\nu,\gamma,\delta,\alpha,\beta\}$ 
correspond to the spin space indices
that ``appear'' in the Hamiltonian on the
bonds that correspond to the pair of 3-site objects 
$\{i,j,k\}$ and $\{l,m,n\}$. To make explicit
what we mean by ``appear'', take the case of
$\sigma^\mu_i \sigma^\nu_j \sigma^\gamma_k$.
If the bonds $\{\langle i,j\rangle,\langle j,k\rangle\}$ 
host $\{\sigma^x_i \sigma^x_j, \sigma^z_j \sigma^z_k\}$,
then $\sigma^\mu_i \sigma^\nu_j \sigma^\gamma_k$ must
equal $\sigma^x_i \sigma^x_j \sigma^z_j \sigma^x_k
= - i \sigma^x_i \sigma^y_j \sigma^x_k $
 for any 
correlator involving this 3-site operator to be non-zero.
If the bonds $\{\langle i,j\rangle,\langle j,k\rangle\}$ 
host $\{\sigma^x_i \sigma^x_j, \sigma^x_j \sigma^x_k\}$,
then $\sigma^\mu_i \sigma^\nu_j \sigma^\gamma_k$ must
equal  
$\sigma^x_i \sigma^x_j \sigma^x_j \sigma^x_k
= \sigma^x_i \sigma^0_j \sigma^x_k
= \sigma^x_i \sigma^x_k$ which
is actually a 2-spin operator for any 
correlator involving this ``3-site'' operator to be non-zero.
And so on. A similar multiplication rule can be
followed to arrive at higher-spin operators with
non-zero correlations. The above structure is 
intuitive as well since it accords with the 
short-range correlations that the Hamiltonian terms 
are trying to favor in the system.
E.g. for the 2-spin operators, it is precisely the
bond energies which have non-zero
correlations, etc.

All of the above proofs can be extended to dynamical
correlators as well. In the dynamical case, 
e.g. for 2-point correlators which is sufficient to illustrate
the main point, we have
$\langle \sigma^\mu_i(t) \sigma^\nu_j(0) \rangle =
\langle e^{- i H t} \sigma^\mu_i e^{i H t} \sigma^\nu_j \rangle$.
We recall here that the ground state expectation value is again
an equal mixture over the ground state manifold
(Eq.~\ref{eq:gsexpectation}).
Since 1) the subdivision of the ground state manifold 
into disjoint sets is based on the conserved quantities 
of the Hamiltonian, and 2) the conserved quantities 
will commute across the $e^{- i H t}$ and $e^{i H t}$ 
factors by definition, the cancellation argument will 
also apply to the dynamical case as well. 
Table~\ref{tab:1} will essentially be reproduced 
for finite $t$ after replacing 
``$\sigma^\mu_{i_\times} \sigma^\nu_{i_\times}$'' with 
``$\sigma^\mu_{i_\times}(t) \sigma^\nu_{i_\times}(0)$'' 
everywhere. Thus the sum over the ground state 
manifold in each disjoint set contributing towards 
$\sigma^\mu_{i_\times}(t) \sigma^\nu_{i_\times}(0)$
 will again cancel out to zero. Similarly, 
 all the arguments in the paper for  
 static multi-spin correlators extend to  
 corresponding dynamical multi-spin correlators as well.

\begin{table*}[t]
\caption{A generic set $A$ of 4 ground states that form
 an equivalence class given two sites $\partial i$ and
 $\partial j$ using the conserved $\{\sigma^z_{\partial k}\}$
 basis for the model of Eq.~\ref{eq:H1} and
 Fig.~\ref{fig:examples1}b.
 \label{tab:2}}
\begin{tabular}{||c | c | c | c||} 
 \hline
 \hline
 $|\psi\rangle$ & $\langle \psi | \sigma^z_{\partial i} 
 | \psi \rangle$ & 
$\langle \psi | \sigma^z_{\partial j} | \psi \rangle$ & 
$\langle \psi | \sigma^\mu_{\partial i} \sigma^\nu_{\partial j} | \psi \rangle$ \\ [0.5ex] 
 \hline\hline
 $|\psi^A_{\text{gs}}\rangle$ & +1 & +1 & $\langle \psi^A_{\text{gs}} |  \sigma^\mu_{\partial i} \sigma^\nu_{\partial j} | \psi^A_{\text{gs}} \rangle$\\ 
 \hline
$\sigma^x_{i} \sigma^x_{\partial i} |\psi^A_{\text{gs}}\rangle$ & -1 & +1 & $\left(2 \delta_{\mu x} - 1\right) \langle  \psi^A_{\text{gs}} |  \sigma^\mu_{\partial i} \sigma^\nu_{\partial j} | \psi^A_{\text{gs}} \rangle$ \\ 
 \hline
 $ \sigma^x_{j} \sigma^x_{\partial j} |\psi^A_{\text{gs}}\rangle$ & +1 & -1 & $ \left(2\delta_{\nu x} - 1\right) \langle  \psi^A_{\text{gs}} |  \sigma^\mu_{\partial i} \sigma^\nu_{\partial j} | \psi^A_{\text{gs}} \rangle$ \\
 \hline
 $\sigma^x_{i} \sigma^x_{\partial i} 
\sigma^x_{j} \sigma^x_{\partial j}
 |\psi^A_{\text{gs}}\rangle$ & -1 & -1 & $\left(2\delta_{\mu x} -1\right)
\left(2\delta_{\nu x} - 1\right) \langle  \psi^A_{\text{gs}} |  \sigma^\mu_{\partial i} \sigma^\nu_{\partial j} | \psi^A_{\text{gs}} \rangle$ \\ [1ex] 
 \hline
\end{tabular}
\end{table*}

\begin{table*}[t]
 \caption{A generic set $A$ of 4 ground states that form
 an equivalence class given two sites $\partial i$ and
 $\partial j$ using the conserved $\{\sigma^x_{\partial k}
 \sigma^x_{\partial k}\}$
 basis for the model of Eq.~\ref{eq:H1} and
 Fig.~\ref{fig:examples1}b.
  \label{tab:3}}
\begin{tabular}{||c | c | c | c||} 
 \hline
 \hline
 $|\psi\rangle$ & $\langle \psi | \sigma^x_i \sigma^x_{\partial i} 
 | \psi \rangle$ & 
$\langle \psi | \sigma^x_j \sigma^x_{\partial j} | \psi \rangle$ & 
$\langle \psi | \sigma^\mu_{\partial i} \sigma^\nu_{\partial j} | \psi \rangle$ \\ [0.5ex] 
 \hline\hline
 $|\psi^A_{\text{gs}}\rangle$ & +1 & +1 & $\langle \psi^A_{\text{gs}} |  \sigma^\mu_{\partial i} \sigma^\nu_{\partial j} | \psi^A_{\text{gs}} \rangle$\\ 
 \hline
$ \sigma^z_{\partial i} |\psi^A_{\text{gs}}\rangle$ & -1 & +1 & $\left(2 \delta_{\mu z} - 1\right) \langle  \psi^A_{\text{gs}} |  \sigma^\mu_{\partial i} \sigma^\nu_{\partial j} | \psi^A_{\text{gs}} \rangle$ \\ 
 \hline
 $ \sigma^z_{\partial j} |\psi^A_{\text{gs}}\rangle$ & +1 & -1 & $ \left(2\delta_{\nu z} - 1\right) \langle  \psi^A_{\text{gs}} |  \sigma^\mu_{\partial i} \sigma^\nu_{\partial j} | \psi^A_{\text{gs}} \rangle$ \\
 \hline
 $\sigma^z_{\partial i} \sigma^z_{\partial j} 
 |\psi^A_{\text{gs}}\rangle$ & -1 & -1 & $\left(2\delta_{\mu z} -1\right)
\left(2\delta_{\nu z} - 1\right) \langle  \psi^A_{\text{gs}} |  \sigma^\mu_{\partial i} \sigma^\nu_{\partial j} | \psi^A_{\text{gs}} \rangle$ \\ [1ex] 
 \hline
\end{tabular}
\end{table*}

Finally, we end this section by considering how
the above arguments apply to the one-dimensional models
of Sec.~\ref{sec:1d} with the representative example
of Eq.~\ref{eq:H1}. Even though there is an extensive
degeneracy in this case, the conserved quantities
$\sigma^x_i \sigma^x_{\partial i}$ can not make 
the ground state expectation 
$\langle \sigma^x_i \sigma^x_j \rangle$
vanish. All other ground state expectation 
$\langle \sigma^\mu_i \sigma^\nu_j \rangle$ 
with $\mu \neq x$
or $\nu \neq x$ do vanish by the above kind of arguments.
We can however use the above kind of arguments to prove
the (classical) spin liquidity on the auxiliary partner
sites. Working in the conserved 
$\{\sigma^z_{\partial i}\}$-basis, we arrive at 
Table~\ref{tab:2},
while working in the conserved 
$\{\sigma^x_i \sigma^x_{\partial i}\}$-basis, we arrive at 
Table~\ref{tab:3}. Combing both of them, 
$\langle  \sigma^\mu_{\partial i} \sigma^\nu_{\partial j}  \rangle = 0$ for any $\mu$, $\nu$ as was already indicated in
our interpretation in Sec.~\ref{sec:1d}.

\subsection{Comparison and contrast with known 
quantum spin liquids}
\label{subsec:compare_contrast}

Since the only non-zero mean-fields are 
those that correspond to the short-range correlations
induced by the Hamiltonian, this suggests the 
absence of other kinds of
non-magnetic spontaneous symmetry breaking as well.
Long-range magnetic correlations are not present as
we have seen earlier with 2-point correlation arguments
for 1-spin operators. This is emblematic of 
a quantum spin liquid. The exact nature of
the spin liquids represented by the models of
Eq.~\ref{eq:2dmodela} and Eq.~\ref{eq:2dmodel} is 
not clear due to a lack of an exact solution.
Certainly a Kitaev or Jordan-Wigner like 
free-fermionization is not operative here. 
We can also already say that they are different
than Kitaev spin liquids due to the extensive
entropy and the anticommutation structure.

It remains
to be seen whether this spin liquid is gapped or 
gapless
modulo the extensive zero modes, even though
the above already
implies that spin correlations are extremely
short-ranged, since there can be fractionalized
excitations in this model analogous to Kitaev
honeycomb model. The absence of other symmetry breaking
orders further suggests fractionalization.
This is an open question. 
We conjecture 
that the spectrum of Eq.~\ref{eq:2dmodela} and Eq.~\ref{eq:2dmodel} 
is gapless when all couplings 
are equal ($J_x=J_z$) and gapped otherwise modulo the
extensive zero modes. A solvable one-dimensional example 
in the form of the chain limit of the models discussed
above is given in Sec.~\ref{subsec:open_questions}
to support this conjecture.

The degeneracies of these models
violate the bound on degeneracy of homogeneous
topological order~\cite{Haah_2021} 
possibly signaling an interpretation of
the ground state manifold in terms
of fractionalized zero modes and, more
generally, gapless fractionalized excitations~\cite{slow_mode}.
Could it be that these models entirely evade a
quasiparticle description to accord with 
the extensive ground state entropy similar to
the fermionic SYK model~\cite{SYK_review}?
Here it may be remarked that in Kitaev honeycomb
model which admits a Majorana
quasiparticle description, 
1-spin correlations (2-point correlators) have the hyperlocal property as a consequence of fractionalization of the spins into Majorana and $Z_2$-gauge degrees of freedom~\cite{Baskaran_Mandal_Shankar_PRL_2007},
long-distance 2-spin correlations (4-point correlators)
are non-zero, decaying exponentially in the gapped 
phase and algebraically in the gapless phase.
The hyperlocal nature of 2-point correlators is also
true for models with the \guillemotleft anticommuting\guillemotright~structure as proved in the previous
section (Sec.~\ref{subsec:liquidity}). The higher multi-spin correlators that
survive the cancellations from the the \guillemotleft anticommuting\guillemotright~mechanism may have similar long distance properties as the Kitaev honecomb model.

There may be additional $O(1)$ degeneracies~\cite{Saptarshi_private}
that we did not touch upon in the previous Sec~\ref{subsec:counting}, i.e. the degeneracies stipulated by the \guillemotleft anticommuting\guillemotright~mechanism is a lower bound on the spectral degeneracy count. 
They may come due to the existence of non-local string operators which remain conserved akin to the global $Z_2$ parity of the TFQIM (and also Eq.~\ref{eq:2dmodel}). 
E.g. for Eq.~\ref{eq:2dmodela} or Eq.~\ref{eq:2dmodelb}, they may be products of the string operator $\prod_{r_y} \sigma^\mu_{(r_x,r_y)}$ on one or more vertical column(s) of sites, and similarly products of $\prod_{r_x} \sigma^\nu_{(r_x,r_y)}$
on one or more horizontal row(s) of sites for appropriate choices of $\mu,\nu$. 
They may also be movable from ``row(s)/column(s) to another row(s)/column(s)'' by the multiplication of appropriate local plaquette conserved quantities without changing the conserved value of the non-local string operators.
Generally such a structure leads to global parity ``superselection'' sectors.
As an example, they could be $\left(\prod_{r_y} \sigma^\mu_{(r_x,r_y)},
\prod_{r_x} \sigma^\mu_{(r_x,r_y)} \right)$ for $\mu \in \{x,y,z\}$. 
Another choice could be $\left(\prod_{r_y} \sigma^x_{(r_x,r_y)}, \prod_{r'_y} \sigma^z_{(r_x,r'_y)}\right)$ where $r_y$ may or may not equal $r'_y$. 
The particular details will depend on the Hamiltonian, e.g. the form of the conserved non-local string  operators may not be same for Eq.~\ref{eq:2dmodela},
~\ref{eq:2dmodelaprime} and ~\ref{eq:2dmodel}.
Such a structure is also present in Kitaev honeycomb model~\cite{Mandal_Shankar_Baskaran_2012}.
Within each global parity superselection sector, there will be an extensive entropy due to the anticommuting mechanism.
Even though this smells like a possible source for topological degeneracy (as the thermodynamic limit is approached similar to TFQIM), it is not clear if the ground state manifold  should be considered topologically ordered in the sense of the Kitaev toric code (which obeys Haah's bound on homogeneous topological order~\cite{Haah_2021}) due to the exponentially large ground states in each global superselection sector. 

Another point of contrast with Kitaev toric code 
is that due to the
mutually conserved nature of the terms in the Kitaev
toric code, they can be used to work out the 
(``$e$'' and ``$m$'') excitations (with mutual anyonic
phase of $\pi$). This allows to explicitly see the 
non-local operations that can be done -- creating
a pair of excitations and annihilating them after taking
one of them across a non-trivial loop on the torus --
to change the topological sector~\cite{Kitaev_2003}. 
In our case, 
the nature of the non-local operations needed to change
the superselection sector is not clear. 
Note that the using the local conserved quantities
of Eq.~\ref{eq:2dmodela} do not help in this since they
also commute with the string operators and do not change
their values. Rather they give rise to the extensive 
entropy in each superselection sector through the
anticommuting mechanism.

For Eq.~\ref{eq:2dmodelb} and
Fig.~\ref{fig:2dmodelb}, we do not have an extensive
degeneracy, but rather a double degeneracy. This is
to be thought of as being similar to the double degeneracy
of TFQIM on the Ising ordered side. Thus this model
is rather Ising ordered with the order being in 
$x$-direction or $z$-direction depending on the
relative magnitudes of $J_x$ and $J_z$ which is
intuitive as well. Note there
is a sub-extensive degeneracy at the transition point
$J_x = J_z$~\cite{Dorier_Becca_Mila_2005}, 
which is indicative of spin liquidity at this
point. Even though free fermionization is not operative
for this model, one can make a mean-field argument
for a Majorana liquid at the quantum phase transition.
One can do a Jordan-Wigner transformation of 
Eq.~\ref{eq:2dmodelb} using a snake-like Jordan-Wigner
string~\cite{JWpreprint} to arrive at
\begin{align}
    H = & \;\; J_x \sum_{\langle \mathbf{r},\mathbf{r}+\mathbf{e_x} \rangle} c^\dagger_{\mathbf{r}} c_{\mathbf{r}+\mathbf{e_x}} +
    c^\dagger_{\mathbf{r}} c^\dagger_{\mathbf{r}+\mathbf{e_x}} + \text{ h.c. }
    + \nonumber \\
    & J_z \sum_{\langle \mathbf{r},\mathbf{r}+\mathbf{e_z} \rangle} \left(n_{\mathbf{r}} - \frac{1}{2} \right)
    \left(n_{\mathbf{r}+\mathbf{e_z}} - \frac{1}{2} \right)
\end{align}
Performing a mean-field decoupling of the four-fermion
term and assuming zero Ising magnetization at the transition,
one arrives at
\begin{equation}
    H_{\text{mf}} = J \sum_{\mathbf{r}}
    c^\dagger_{\mathbf{r}} c_{\mathbf{r}+\mathbf{e_x}} +
    c^\dagger_{\mathbf{r}} c^\dagger_{\mathbf{r}+\mathbf{e_x}}
    + c^\dagger_{\mathbf{r}} c_{\mathbf{r}+\mathbf{e_z}} +
    c^\dagger_{\mathbf{r}} c^\dagger_{\mathbf{r}+\mathbf{e_z}} + \text{ h.c. }
\end{equation}
where $J = J_x = J_z$. This is a $p$-wave 
superconductor of spinless fermions with gapless
nodes in the two dimensional Brillouin zone 
with Majorana excitations,
very analogous to the TFQIM transition in one dimension.

\section{Conclusion}
\label{sec:conclu}

\subsection{Summary of results}
\label{subsec:summary}

This work describes a construction for spin-$\frac{1}{2}$ models which results in an extensive residual ground state entropy and quantum spin liquidity. 
For any Hamiltonian, if it hosts an \guillemotleft anticommuting\guillemotright~algebra of local conserved quantities that have extensive cardinality, such behaviour
would manifest. 
The discussion around Eq.~\ref{eq:2dmodela} and Eq.~\ref{eq:2dmodel} in Sec.~\ref{sec:2d} gives examples of this \guillemotleft anticommuting\guillemotright~structure
or mechanism.
The extensive entropy property and the concomitant spin liquidity holds throughout the full eigenspectrum.
This construction is natural in the presence of bond-dependent couplings and the resultant physics is exposed through the examples of the models in (Eq.~\ref{eq:2dmodela}, Fig~\ref{fig:2dmodela}) and (Eq.~\ref{eq:2dmodel}, Fig~\ref{fig:2dmodel}).

The basic aspects of extensive entropy was exposed
through pedagogical one-dimensional examples in Sec.~\ref{sec:1d}.
The main example of Eq.~\ref{eq:H1} as a variant of 
the well-known transverse field quantum Ising model
had effectively classical conserved (auxiliary) spins
which did not void quantum (Ising) order. In Sec.~\ref{sec:2d},
we constructed models without any effective classical
degrees of freedom. This led to provable quantum spin 
liquidity (Sec.~\ref{subsec:liquidity}) apart from
the extensive residual entropy (Sec.~\ref{subsec:counting}).
The hyperlocal vanishing of the static and dynamical spin 2-point correlators is a natural consequence of the local \guillemotleft anticommutation\guillemotright~structure and the resulting spectral degeneracies. This is in fact true throughout the spectrum and thus for all temperatures. A comparison and contrast
of these quantum spin liquids to other known quantum spin
liquids and models was given in Sec.~\ref{subsec:compare_contrast}.

\begin{figure}[t]
    \centering
    \includegraphics[width=0.65\linewidth]{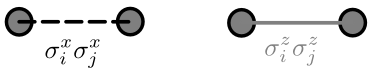} \\
    \includegraphics[width=0.7\linewidth]{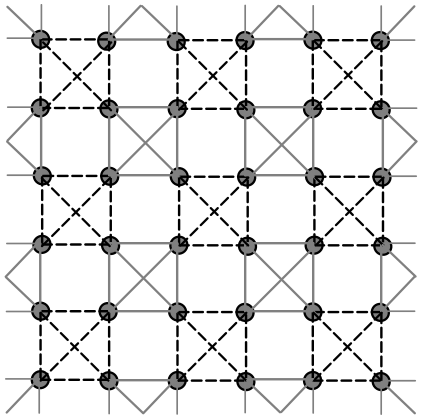}
    \caption{A model on the two-dimensional analog 
    of the three-dimensional pyrochlore lattice 
    with extensive ground state entropy enforced
    by the anticommuting mechanism. Analogous 
    results from Sec.~\ref{sec:2d}
    will apply to this model as well.}
    \label{fig:2dpyrocholore}
\end{figure}

Another aspect is the gauge-like nature of the extensively many \guillemotleft anticommuting\guillemotright~local conserved quantities in these constructions.
This may be a novel way in which gauge-like physical degrees of freedom emerge in quantum spin-$\frac{1}{2}$ systems, e.g. when comparing to the Levin-Wen model and Kitaev's toric code, Kitaev's honeycomb model, Haah's code and X-cube model ~\cite{Kitaev_2003,Levin_Wen_2005,Haah_2011,Vijay_Haah_Fu_2016,Pretko_Chen_You_2020} all of which have only commuting conserved sets.
This anticommuting or non-commuting mechanism can in general operate in any number of dimensions.
A natural construction is similar to the model of Eq.~\ref{eq:2dmodela} on the three-dimensional pyrochlore lattice. 
A two-dimensional version of this on the ``two-dimensional pyrochlore lattice'' is shown in Fig.~\ref{fig:2dpyrocholore}.
Also may be mentioned that the anticommuting mechanism seems natural for spin models, in particular spin-$\frac{1}{2}$, since bosons do not naturally accommodate anticommutation and it does not appear so for fermions as well~\cite{footnote_SYK}.
Higher spin models can accommodate more general forms of non-commutation and local versions of non-commutation beyond anticommutation will be interesting to find.

\subsection{Remarks on physical consequences and
realization}
\label{subsec:physical_realization}

The physical consequences originating from the
\guillemotleft anticommuting\guillemotright~mechanism 
in a particular two-dimensional model
not discussed here and
closely related to case (a) in Sec.~\ref{sec:1d}
has been discussed in Ref.~\cite{JWpreprint}.
In general, we expect a residual ground state entropy
even when the excitation spectrum is gapped due to
the degeneracy of the ground state manifold. 
This can lead to a two-peak structure in the magnetic
component of the specific heat similar to Kitaev spin
liquids~\cite{JWpreprint}.
The hyperlocal nature of 2-point spin correlators
proved in Sec.~\ref{subsec:liquidity}
for the higher-dimensional quantum constructions 
of Sec.~\ref{sec:2d}  
implies no magnetic ordering and completely featureless
spin structure factors at all energies. Spin structure
factors of quantum magnets are commonly probed 
in inelastic neutron scattering experiments.
The presence of global superselection sectors discussed in
Sec.~\ref{subsec:compare_contrast} leaves open the
intriguing prospect of topological order coexisting with
quantum spin liquidity.

In terms of experimental realizations, artificial quantum systems
seem to be the best bet for observing the physics of these models. 
One such example is Ref.~\cite{Homeier_etal_2021} which proposes
a superconducting qubit based analog simulation of bond-dependent
lattice spin model physics.
In a solid state material context, perhaps some quantum compass
type lattice realization~\cite{Kugel_Khomskii_1973,Kugel_Khomskii_1982} may accommodate the needed structure of the couplings as an alternative to the Jackeli-Khaillulin mechanism~\cite{Jackeli_Khaliullin_PRL_2009} for Kitaev type bond-dependent couplings. 
Another possibility is that the basic physics of these models can be made operational in the low energy physics of some quantum spin-$\frac{1}{2}$ ice materials~\cite{spin_ice_review} if one arranges for the quantization axes of the quantum Ising couplings to ``stagger" across the lattice in some sensible way. 
Quantum fluctuations in quantum spin ices are usually induced through a transverse field term as opposed to the \guillemotleft anticommuting\guillemotright~mechanism. 
A Rydberg-atom based artificial implementation of quantum spin ices has recently been proposed recently in Ref.~\cite{Shah_etal_2025} which can provide another direction for an experimental realization of these models.

\subsection{Outlook and open questions}
\label{subsec:open_questions}


\begin{enumerate}
    \item A detail-oriented issue is if there
    are there additional ``accidental'' degeneracies at
    fine-tuned ratios of the couplings
    apart from the generic exponential degeneracies 
    in the spirit of Ref.~\cite{Dorier_Becca_Mila_2005}.

    \item A general question is regarding the nature
    of excitations in these models. Is there a finite
    energy gap above the ground state manifold or not?
    Are there 
    fractionalized excitations in models with 
    the anticommutation structure and concomitant
    extensive entropies such as the ones discussed
    in this paper? Could they entirely evade
    a quasiparticle description?

    One example where we can explicitly work out the 
    nature of the excitations is a one-dimensional
    Kitaev chain model that can be considered a cousin of either 
    the Kitaev honeycomb model or the models in 
    Eqs.~\ref{eq:2dmodela} and ~\ref{eq:2dmodelaprime}, i.e.
    \begin{equation}
        H = J_x \sum_{\textbf{---}_x} \sum_{\langle i,j \rangle \in \; \textbf{---}_x} 
            \sigma^x_i \sigma^x_j 
            +
            J_z \sum_{\textbf{---}_z} \sum_{\langle i,j \rangle \in \; \textbf{---}_z} 
            \sigma^z_i \sigma^z_j 
    \end{equation}
    The above model has received attention in the literature
    ~\cite{Agrapidis_Brink_Nishimoto_2018,Morris_etal_2021} in
    the context of Kitaev spin liquid physics.
    The \guillemotleft anticommuting \guillemotright~sets of local conserved quantities are
    \begin{itemize}
        \item $\sigma^z_i \sigma^z_j$ on $\textbf{---}_x$
        \item $\sigma^x_i \sigma^x_j$ on $\textbf{---}_z$
    \end{itemize}
    which again guarantees an extensive ground state entropy.
    The above Hamiltonian can also be reduced to a fermionic
    quadratic form using Majorana operators~\cite{Fu_Knolle_Perkins_PRB_2018}
    to yield
    \begin{equation}
        H = J_x \sum_{\textbf{---}_x} \sum_{\langle i,j \rangle \in \; \textbf{---}_x} 
           u^x_{ij} \; \gamma_i \gamma_j 
            +
            J_z \sum_{\textbf{---}_z} \sum_{\langle i,j \rangle \in \; \textbf{---}_z} 
             u^z_{ij} \; \gamma_i \gamma_j 
    \end{equation}
    where $\gamma^\dagger = \gamma$ and $u^\mu_{ij}$ are the
    $Z_2$-conserved quantities in the Kitaev representation
    ($\propto \gamma^\mu_i \gamma^\mu_j$), whose spectrum can be
    explicitly worked out (Eq.~32 of Ref.~\cite{Kitaev_2006}
    with $J_y=0$).
    It is nothing but an edge of the famous triangle phase
    diagram of the Kitaev honeycomb model (Fig.~5 of Ref.~\cite{Kitaev_2006}). The excitation spectrum consists
    of Majorana modes that are gapped for $J_x \neq J_z$ and become
    gapless at the quantum phase transition $J_x = J_z$. Note the
    two phases on \emph{both} sides have the same topological order
    describable by the toric code. Furthermore, each mode is 
    necessarily extensively degenerate even when gapped. This is thus
    a violation of Haah's bound~\cite{Haah_2021} for gapped topological
    order. Whether this is more than a pathology remains to be seen.
    However this is certainly natural when seen as a one-dimensional
    limit of Eq.~\ref{eq:2dmodela}. The extensive ground
    state entropy can then be interpreted as 
    ``physicalizing''
    ($Z_2$) gauge artifacts of an infinitely long one dimensional chain!~\cite{JWpreprint}.

    In light of the above discussion, where we could pinpoint the
    nature of the excitations in the solvable one-dimensional case,
    one may wonder how can these models evade a quasiparticle 
    description as conjectured before. This may still be the 
    case in terms of the dynamical correlations. 
    Even though the spectrum has a Majorana quadratic form in 
    each block of Hamiltonian in the above example, time
    evolution may ``instantaneously'' start mixing the many blocks for \emph{generic} initial states, which
    may effectively render a quasiparticle description ineffective
    for dynamical correlations.

    \item In presence of deformations that void
    the anticommutation structure, what would be
    a generic consequence in models close to the
    limit where the anticommuting structure remains intact? 
    An example of this was seen
    in Sec.~\ref{subsec:1dextensions} where the deformation
    was one of the set of conserved quantities. If the
    deformation is not one of the 
    conserved quantities, does that generically imply the 
    appearance of slow modes made out of the degenerate
    manifold as conjectured in Ref.~\cite{JWpreprint}, 
    similar to what happens at the sub-extensively
    degenerate quantum phase transition in the 
    $90^\circ$ compass model~\cite{Dorier_Becca_Mila_2005}

    \item The  entanglement structure in the ground
    state manifold is certainly worth investigating. Can
    there be a way to make progress using the anticommutation
    structure without knowing the exact ground state
    solutions?

    \item Numerics will be useful to investigate the open questions in the points one to four written above. The thermodynamic properties of a particular case of Eq.~\ref{eq:2dmodela} with translational invariance within the broader class of quantum spin liquids dilineated in this work was numerically studied by Wenzel and Janke~\cite{Wenzel_Janke_2009}. Ref.~\cite{Wenzel_Janke_2008} dubbed this case as the plaquette orbital model motivated by considerations related to compass model physics~\cite{Wenzel_Janke_2008,Wenzel_Janke_Laeuchli_2010}.  Thus it concluded a lower bound of two on the degeneracy. See the discussion below Eq.~4 of Ref.~\cite{Wenzel_Janke_2009} where it says, ``This shows that every energy eigenvalue of $H_{\text{POM}}$ [plaquette orbital model] is at least twofold degenerate. Performing an exact diagonalization \ldots~confirming this conclusion. The POM [plaquette orbital model] hence possesses the same behavior as the CM [compass model] in this regard.'' The extensively larger lower bound (of $2^{\#\text{unit cells}}$) on the degeneracy and the consequent provable spin liquidity was not reported presumably because the lens was that of the existing compass model results. However  it did find the absence of spin or magnetic ordering numerically at finite temperatures.  This finite-temperature spin liquidity is also to be expected given the spin liquidity arguments laid out in Sec.~\ref{subsec:liquidity}. Since these arguments are purely in terms of the \guillemotleft anticommuting\guillemotright~algebraic structure, they are therefore true for the full eigenspectrum and thus at all temperatures. A follow-up studied the classical version of the plaquette orbital model in more detail~\cite{Biskup_Kotecky_2010}.

    \item Could there be other physical interrelations
    between the ground states more than what the 
    anticommutation structure stipulates, or at least a
    more detailed view of them? A classical 
    example of this would be from constrained statistical
    physics models, e.g. the relation between the different classical spin ice ground states as being connected
    by loops where the spin orientations are flipped to
    connect them.
    Without the 
    knowledge of the exact ground states, this
    is not obvious. 
    
    \item Of course, all of the above motivates
    constructing solvable cousins of these models.
    Constructions which are solvable and have
extensive entropy can be written
down, but it is not evident 
how to avoid solvable constructions which do
not have any effective classical variables
(conserved $\sigma^\mu_i$ for some $\mu$ and 
subset of sites).
One such construction has been discussed in
Ref.~\cite{JWpreprint}. Constructions that do
not have
any such effective classical degrees of freedom,
host a generically extensive ground state entropy through
the \guillemotleft anticommuting\guillemotright~mechanism 
as the models discussed in this work, and are solvable through 
some means
would be very interesting to study and is an open
question.

    \item At a framework level, this work suggests a 
    general theory
for constructing models with extensive ground state
entropy in the spirit of what Refs.~\cite{Ogura_etal_2020,Chapman_Flammia_2020} 
and related papers~\cite{Minami_2016,Minami_2017,Minami_Yanagihara_2020,
Chapman_Elman_Flammia_2021,Chapman_Elman_Mann_2023,Fendley_Pozsgay_2024}
do for spin models with free fermion spectra~\cite{Fendley_2019}. 

    \item Finally it is not fully clear how does the
strongly correlated physics described here fit in
the atlas of strongly correlated physics such as 
many-body topological orders and/or 
the absence of quasiparticle descriptions.
\end{enumerate}

\subsection{Final remarks}
\label{subsec:speculations}

 The gauge-like aspect of these models mentioned before
 in Sec.~\ref{subsec:summary}
deserves more exploration it seems to the author. 
For example, the interpretation of the extensive ground
    state entropy in the one dimensional Kitaev chain
    model as ``physicalizing"
    ($Z_2$) gauge artifacts in an infinitely long one dimensional chain alluded to in the second point of Sec.~\ref{subsec:open_questions}; does this interpretation
    somehow extend to higher dimensions?
        A related quasi-one-dimensional 
        example involving thin strips in the existing
literature is Ref.~\cite{Chen_etal_2024} where the
authors discuss the extensive entropy generation to be 
related to the recent developments under the rubric of 
``higher-form'' symmetries~\cite{higher_form_footnote}.
One difference is that non-commuting conserved operators 
are on system-width spanning long strings in the model of
Ref.~\cite{Chen_etal_2024}, whereas in the constructions
discussed here, the non-commuting quantities are local 
throughout in this sense with support over $O(1)$ lattice
sites. It remains to be seen if there
is a higher-form symmetry perspective on the 
\guillemotleft anticommuting\guillemotright~mechanism
that may inform further on this issue.

Another question is if there exists a field-theoretic formulation of these
    models in the continuum analogous to 
    Chern-Simons field theories for
    many-body topological orders? A remark here would be
    that Landau levels also have an extensive degeneracy,
    however proportional
    to the system area and, unlike this paper, 
    not exponentially large. 
    Chern-Simons theories give a field-theoretic understanding
    of the resultant many-body insulating states with
    topological order~\cite{Tong_notes_2016}. 
    However time reversal is explicitly broken in these
    cases due to
    the presence of a magnetic field. In all the models
    discussed in this work, time reversal is preserved
    since the Hamiltonians are composed of 2-spin terms
    and $\sigma^\mu_i \rightarrow -\sigma^\mu_i$ under
    time reversal for spin-$\frac{1}{2}$ degrees of 
    freedom. Furthermore, from an algebraic point of
    view, the generation of degeneracy in Landau levels is
    due to large symmetry or quantum number generators which
    do not appear in the Hamiltonian which is different
    from the anticommuting mechanism. For the SYK model
    which can have an extensive ground state entropy, there 
    exist dynamical (mean-)field theory formulations
    in the continuum~\cite{SYK_review}.

The final physical point that is perhaps of relevance 
relates to quantum chaos bounds~\cite{Maldacena_Shenker_Stanford_2016}.
It has been shown that the SYK model saturates
this bound~\cite{Kitaev_talk_2015,Kitaev_Suh_2018,
Maldacena_Stanford_2016}. Given the extensive ground
state entropy of the SYK model~\cite{footnote_SYK_review_secVD}
and the relation of zero modes to the saturation of the
chaos bound~\cite{Maldacena_Stanford_2016, footnote_SYK_review_secXIIE}, it is 
tempting to conjecture that the spin models discussed here
may also approach -- perhaps saturate -- the quantum chaos bound
like the original Sachdev-Ye model~\cite{Sachdev_Ye_1993}.
These models and in particular the two-dimensional ones 
of Eq.~\ref{eq:2dmodela}/Fig.~\ref{fig:2dmodel} and
Eq.~\ref{eq:2dmodel}/Fig.~\ref{fig:2dmodel}
may then provide spin models with
\emph{local} interactions that approach the quantum chaos
bound with appropriate butterfly velocities~\cite{OTOC_encyclopedia}. In this context, the speculation of the absence
of a quasiparticle description made earlier in
Sec.~\ref{subsec:liquidity} may also 
be pertinent~\cite{black_hole}.

\SPhide{
Another speculation would then be if these
models with spin-$\frac{1}{2}$ microscopic degrees of
freedom or qubits connect somehow to black hole physics in 
analogy with the connection between the SYK model and
charged black holes~\cite{Sachdev_2010,Kitaev_talk_2015}.
Could these models have something to say about
quantum gravity analogous to the SYK models?
}

\section*{Acknowledgements}
Discussions with Ajit Balram, Sumilan Banerjee, Arkya Chatterjee, Pieter Claeys, Kedar Damle, Prashant Kumar, Abhijit Gadde, Darshan Joshi, Saptarshi Mandal, Shiraz Minwalla, Roderich Moessner, Onkar Parrikar, Karlo Penc, Rajdeep Sensarma and Sandip Trivedi are gratefully acknowledged.
Funding support from SERB-DST, India (superseded by ANRF-DST established through an Act of Parliament: ANRF Act, 2023) via Grant No. MTR/2022/000386 is acknowledged.

\bibliography{refs.bib}

\begin{thebibliography}{84}%
\makeatletter
\providecommand \@ifxundefined [1]{%
 \@ifx{#1\undefined}
}%
\providecommand \@ifnum [1]{%
 \ifnum #1\expandafter \@firstoftwo
 \else \expandafter \@secondoftwo
 \fi
}%
\providecommand \@ifx [1]{%
 \ifx #1\expandafter \@firstoftwo
 \else \expandafter \@secondoftwo
 \fi
}%
\providecommand \natexlab [1]{#1}%
\providecommand \enquote  [1]{``#1''}%
\providecommand \bibnamefont  [1]{#1}%
\providecommand \bibfnamefont [1]{#1}%
\providecommand \citenamefont [1]{#1}%
\providecommand \href@noop [0]{\@secondoftwo}%
\providecommand \href [0]{\begingroup \@sanitize@url \@href}%
\providecommand \@href[1]{\@@startlink{#1}\@@href}%
\providecommand \@@href[1]{\endgroup#1\@@endlink}%
\providecommand \@sanitize@url [0]{\catcode `\\12\catcode `\$12\catcode
  `\&12\catcode `\#12\catcode `\^12\catcode `\_12\catcode `\%12\relax}%
\providecommand \@@startlink[1]{}%
\providecommand \@@endlink[0]{}%
\providecommand \url  [0]{\begingroup\@sanitize@url \@url }%
\providecommand \@url [1]{\endgroup\@href {#1}{\urlprefix }}%
\providecommand \urlprefix  [0]{URL }%
\providecommand \Eprint [0]{\href }%
\providecommand \doibase [0]{https://doi.org/}%
\providecommand \selectlanguage [0]{\@gobble}%
\providecommand \bibinfo  [0]{\@secondoftwo}%
\providecommand \bibfield  [0]{\@secondoftwo}%
\providecommand \translation [1]{[#1]}%
\providecommand \BibitemOpen [0]{}%
\providecommand \bibitemStop [0]{}%
\providecommand \bibitemNoStop [0]{.\EOS\space}%
\providecommand \EOS [0]{\spacefactor3000\relax}%
\providecommand \BibitemShut  [1]{\csname bibitem#1\endcsname}%
\let\auto@bib@innerbib\@empty
\bibitem [{\citenamefont {Peierls}(1936)}]{Peierls_1936}%
  \BibitemOpen
  \bibfield  {author} {\bibinfo {author} {\bibfnamefont {R.}~\bibnamefont
  {Peierls}},\ }\bibfield  {title} {\bibinfo {title} {On ising’s model of
  ferromagnetism},\ }\href {https://doi.org/10.1017/S0305004100019174}
  {\bibfield  {journal} {\bibinfo  {journal} {Mathematical Proceedings of the
  Cambridge Philosophical Society}\ }\textbf {\bibinfo {volume} {32}},\
  \bibinfo {pages} {477–481} (\bibinfo {year} {1936})}\BibitemShut {NoStop}%
\bibitem [{\citenamefont {Bonati}(2014)}]{Bonati_2014}%
  \BibitemOpen
  \bibfield  {author} {\bibinfo {author} {\bibfnamefont {C.}~\bibnamefont
  {Bonati}},\ }\bibfield  {title} {\bibinfo {title} {The peierls argument for
  higher dimensional ising models},\ }\href
  {https://doi.org/10.1088/0143-0807/35/3/035002} {\bibfield  {journal}
  {\bibinfo  {journal} {European Journal of Physics}\ }\textbf {\bibinfo
  {volume} {35}},\ \bibinfo {pages} {035002} (\bibinfo {year}
  {2014})}\BibitemShut {NoStop}%
\bibitem [{\citenamefont {Elitzur}(1975)}]{Elitzur_1975}%
  \BibitemOpen
  \bibfield  {author} {\bibinfo {author} {\bibfnamefont {S.}~\bibnamefont
  {Elitzur}},\ }\bibfield  {title} {\bibinfo {title} {Impossibility of
  spontaneously breaking local symmetries},\ }\href
  {https://doi.org/10.1103/PhysRevD.12.3978} {\bibfield  {journal} {\bibinfo
  {journal} {Phys. Rev. D}\ }\textbf {\bibinfo {volume} {12}},\ \bibinfo
  {pages} {3978} (\bibinfo {year} {1975})}\BibitemShut {NoStop}%
\bibitem [{\citenamefont {{Fr{\"o}hlich}}\ \emph {et~al.}(1981)\citenamefont
  {{Fr{\"o}hlich}}, \citenamefont {{Morchio}},\ and\ \citenamefont
  {{Strocchi}}}]{Frohlich_Morchio_Strocchi_1981}%
  \BibitemOpen
  \bibfield  {author} {\bibinfo {author} {\bibfnamefont {J.}~\bibnamefont
  {{Fr{\"o}hlich}}}, \bibinfo {author} {\bibfnamefont {G.}~\bibnamefont
  {{Morchio}}},\ and\ \bibinfo {author} {\bibfnamefont {F.}~\bibnamefont
  {{Strocchi}}},\ }\bibfield  {title} {\bibinfo {title} {{Higgs phenomenon
  without symmetry breaking order parameter}},\ }\href
  {https://doi.org/10.1016/0550-3213(81)90448-X} {\bibfield  {journal}
  {\bibinfo  {journal} {Nuclear Physics B}\ }\textbf {\bibinfo {volume}
  {190}},\ \bibinfo {pages} {553} (\bibinfo {year} {1981})}\BibitemShut
  {NoStop}%
\bibitem [{\citenamefont {Ginzburg}(1961)}]{Ginzburg_1961}%
  \BibitemOpen
  \bibfield  {author} {\bibinfo {author} {\bibfnamefont {V.~L.}\ \bibnamefont
  {Ginzburg}},\ }\bibfield  {title} {\bibinfo {title} {Some remarks on phase
  transitions of the second kind and the microscopic theory of ferroelectric
  materials},\ }\href {https://cir.nii.ac.jp/crid/1572261549733336704}
  {\bibfield  {journal} {\bibinfo  {journal} {Soviet Phys. Solid State}\
  }\textbf {\bibinfo {volume} {2}},\ \bibinfo {pages} {1824} (\bibinfo {year}
  {1961})}\BibitemShut {NoStop}%
\bibitem [{\citenamefont {Harris}(1974)}]{Harris_1974}%
  \BibitemOpen
  \bibfield  {author} {\bibinfo {author} {\bibfnamefont {A.~B.}\ \bibnamefont
  {Harris}},\ }\bibfield  {title} {\bibinfo {title} {Effect of random defects
  on the critical behaviour of ising models},\ }\href
  {https://doi.org/10.1088/0022-3719/7/9/009} {\bibfield  {journal} {\bibinfo
  {journal} {Journal of Physics C: Solid State Physics}\ }\textbf {\bibinfo
  {volume} {7}},\ \bibinfo {pages} {1671} (\bibinfo {year} {1974})}\BibitemShut
  {NoStop}%
\bibitem [{\citenamefont {Vojta}\ and\ \citenamefont
  {Hoyos}(2014)}]{Vojta_Hoyos_2014}%
  \BibitemOpen
  \bibfield  {author} {\bibinfo {author} {\bibfnamefont {T.}~\bibnamefont
  {Vojta}}\ and\ \bibinfo {author} {\bibfnamefont {J.~A.}\ \bibnamefont
  {Hoyos}},\ }\bibfield  {title} {\bibinfo {title} {Criticality and quenched
  disorder: Harris criterion versus rare regions},\ }\href
  {https://doi.org/10.1103/PhysRevLett.112.075702} {\bibfield  {journal}
  {\bibinfo  {journal} {Phys. Rev. Lett.}\ }\textbf {\bibinfo {volume} {112}},\
  \bibinfo {pages} {075702} (\bibinfo {year} {2014})}\BibitemShut {NoStop}%
\bibitem [{\citenamefont {Imry}\ and\ \citenamefont {Ma}(1975)}]{Imry_Ma_1975}%
  \BibitemOpen
  \bibfield  {author} {\bibinfo {author} {\bibfnamefont {Y.}~\bibnamefont
  {Imry}}\ and\ \bibinfo {author} {\bibfnamefont {S.-k.}\ \bibnamefont {Ma}},\
  }\bibfield  {title} {\bibinfo {title} {Random-field instability of the
  ordered state of continuous symmetry},\ }\href
  {https://doi.org/10.1103/PhysRevLett.35.1399} {\bibfield  {journal} {\bibinfo
   {journal} {Phys. Rev. Lett.}\ }\textbf {\bibinfo {volume} {35}},\ \bibinfo
  {pages} {1399} (\bibinfo {year} {1975})}\BibitemShut {NoStop}%
\bibitem [{\citenamefont {Aizenman}\ and\ \citenamefont
  {Wehr}(1989)}]{Aizenman_Wehr_1989}%
  \BibitemOpen
  \bibfield  {author} {\bibinfo {author} {\bibfnamefont {M.}~\bibnamefont
  {Aizenman}}\ and\ \bibinfo {author} {\bibfnamefont {J.}~\bibnamefont
  {Wehr}},\ }\bibfield  {title} {\bibinfo {title} {Rounding of first-order
  phase transitions in systems with quenched disorder},\ }\href
  {https://doi.org/10.1103/PhysRevLett.62.2503} {\bibfield  {journal} {\bibinfo
   {journal} {Phys. Rev. Lett.}\ }\textbf {\bibinfo {volume} {62}},\ \bibinfo
  {pages} {2503} (\bibinfo {year} {1989})}\BibitemShut {NoStop}%
\bibitem [{\citenamefont {Castelnovo}\ \emph {et~al.}(2012)\citenamefont
  {Castelnovo}, \citenamefont {Moessner},\ and\ \citenamefont
  {Sondhi}}]{spin_ice_review}%
  \BibitemOpen
  \bibfield  {author} {\bibinfo {author} {\bibfnamefont {C.}~\bibnamefont
  {Castelnovo}}, \bibinfo {author} {\bibfnamefont {R.}~\bibnamefont
  {Moessner}},\ and\ \bibinfo {author} {\bibfnamefont {S.}~\bibnamefont
  {Sondhi}},\ }\bibfield  {title} {\bibinfo {title} {Spin ice,
  fractionalization, and topological order (sec. 2a)},\ }\href
  {https://doi.org/https://doi.org/10.1146/annurev-conmatphys-020911-125058}
  {\bibfield  {journal} {\bibinfo  {journal} {Annual Review of Condensed Matter
  Physics}\ }\textbf {\bibinfo {volume} {3}},\ \bibinfo {pages} {35} (\bibinfo
  {year} {2012})}\BibitemShut {NoStop}%
\bibitem [{\citenamefont {Chowdhury}\ \emph {et~al.}(2022)\citenamefont
  {Chowdhury}, \citenamefont {Georges}, \citenamefont {Parcollet},\ and\
  \citenamefont {Sachdev}}]{SYK_review}%
  \BibitemOpen
  \bibfield  {author} {\bibinfo {author} {\bibfnamefont {D.}~\bibnamefont
  {Chowdhury}}, \bibinfo {author} {\bibfnamefont {A.}~\bibnamefont {Georges}},
  \bibinfo {author} {\bibfnamefont {O.}~\bibnamefont {Parcollet}},\ and\
  \bibinfo {author} {\bibfnamefont {S.}~\bibnamefont {Sachdev}},\ }\bibfield
  {title} {\bibinfo {title} {Sachdev-ye-kitaev models and beyond: Window into
  non-fermi liquids (sec. 5d)},\ }\href
  {https://doi.org/10.1103/RevModPhys.94.035004} {\bibfield  {journal}
  {\bibinfo  {journal} {Rev. Mod. Phys.}\ }\textbf {\bibinfo {volume} {94}},\
  \bibinfo {pages} {035004} (\bibinfo {year} {2022})}\BibitemShut {NoStop}%
\bibitem [{thi()}]{third_law}%
  \BibitemOpen
  \bibinfo {note} {Http://www.nernst.de/\#theorem, Memorial website of Walther
  Hermann Nernst}\BibitemShut {NoStop}%
\bibitem [{\citenamefont {{Masanes}}\ and\ \citenamefont
  {{Oppenheim}}(2017)}]{Masanes_Oppenheim_2017}%
  \BibitemOpen
  \bibfield  {author} {\bibinfo {author} {\bibfnamefont {L.}~\bibnamefont
  {{Masanes}}}\ and\ \bibinfo {author} {\bibfnamefont {J.}~\bibnamefont
  {{Oppenheim}}},\ }\bibfield  {title} {\bibinfo {title} {{A general derivation
  and quantification of the third law of thermodynamics}},\ }\href
  {https://doi.org/10.1038/ncomms14538} {\bibfield  {journal} {\bibinfo
  {journal} {Nature Communications}\ }\textbf {\bibinfo {volume} {8}},\
  \bibinfo {eid} {14538} (\bibinfo {year} {2017})},\ \Eprint
  {https://arxiv.org/abs/1412.3828} {arXiv:1412.3828 [quant-ph]} \BibitemShut
  {NoStop}%
\bibitem [{\citenamefont {Wigner}(1934)}]{Wigner_1934}%
  \BibitemOpen
  \bibfield  {author} {\bibinfo {author} {\bibfnamefont {E.}~\bibnamefont
  {Wigner}},\ }\bibfield  {title} {\bibinfo {title} {On the interaction of
  electrons in metals},\ }\href {https://doi.org/10.1103/physrev.46.1002}
  {\bibfield  {journal} {\bibinfo  {journal} {Physical Review}\ }\textbf
  {\bibinfo {volume} {46}},\ \bibinfo {pages} {1002–1011} (\bibinfo {year}
  {1934})}\BibitemShut {NoStop}%
\bibitem [{\citenamefont {{Wikipedia contributors}}(2024)}]{Wigner_wiki}%
  \BibitemOpen
  \bibfield  {author} {\bibinfo {author} {\bibnamefont {{Wikipedia
  contributors}}},\ }\href
  {https://en.wikipedia.org/w/index.php?title=Wigner_crystal&oldid=1221946977}
  {\bibinfo {title} {Wigner crystal --- {Wikipedia}{,} the free encyclopedia}}
  (\bibinfo {year} {2024}),\ \bibinfo {note} {[Online; accessed
  26-August-2024]; see the sections on experimental realization and related
  references.}\BibitemShut {Stop}%
\bibitem [{\citenamefont {{Jordan}}\ and\ \citenamefont
  {{Wigner}}(1928)}]{Jordan_Wigner_1928}%
  \BibitemOpen
  \bibfield  {author} {\bibinfo {author} {\bibfnamefont {P.}~\bibnamefont
  {{Jordan}}}\ and\ \bibinfo {author} {\bibfnamefont {E.}~\bibnamefont
  {{Wigner}}},\ }\bibfield  {title} {\bibinfo {title} {{{\"U}ber das Paulische
  {\"A}quivalenzverbot}},\ }\href {https://doi.org/10.1007/BF01331938}
  {\bibfield  {journal} {\bibinfo  {journal} {Zeitschrift fur Physik}\ }\textbf
  {\bibinfo {volume} {47}},\ \bibinfo {pages} {631} (\bibinfo {year}
  {1928})}\BibitemShut {NoStop}%
\bibitem [{\citenamefont {Pujari}(2024)}]{JWpreprint}%
  \BibitemOpen
  \bibfield  {author} {\bibinfo {author} {\bibfnamefont {S.}~\bibnamefont
  {Pujari}},\ }\bibfield  {title} {\bibinfo {title} {A solvable embedding
  mechanism for one-dimensional spinless and majorana fermions in
  higher-dimensional spin-1/2 magnets}\ }\href
  {https://doi.org/10.48550/arXiv.2406.17034} {10.48550/arXiv.2406.17034}
  (\bibinfo {year} {2024}),\ \Eprint {https://arxiv.org/abs/2406.17034}
  {arXiv:2406.17034 [cond-mat.str-el]} \BibitemShut {NoStop}%
\bibitem [{\citenamefont {Nussinov}\ and\ \citenamefont
  {Ortiz}(2008)}]{Nussinov_Ortiz_2008}%
  \BibitemOpen
  \bibfield  {author} {\bibinfo {author} {\bibfnamefont {Z.}~\bibnamefont
  {Nussinov}}\ and\ \bibinfo {author} {\bibfnamefont {G.}~\bibnamefont
  {Ortiz}},\ }\bibfield  {title} {\bibinfo {title} {Orbital order driven
  quantum criticality},\ }\href {https://doi.org/10.1209/0295-5075/84/36005}
  {\bibfield  {journal} {\bibinfo  {journal} {EPL (Europhysics Letters)}\
  }\textbf {\bibinfo {volume} {84}},\ \bibinfo {pages} {36005} (\bibinfo {year}
  {2008})}\BibitemShut {NoStop}%
\bibitem [{\citenamefont {Nussinov}\ and\ \citenamefont
  {Ortiz}(2009)}]{Nussinov_Ortiz_2009}%
  \BibitemOpen
  \bibfield  {author} {\bibinfo {author} {\bibfnamefont {Z.}~\bibnamefont
  {Nussinov}}\ and\ \bibinfo {author} {\bibfnamefont {G.}~\bibnamefont
  {Ortiz}},\ }\bibfield  {title} {\bibinfo {title} {Bond algebras and exact
  solvability of hamiltonians: Spin $s=\frac{1}{2}$ multilayer systems},\
  }\href {https://doi.org/10.1103/PhysRevB.79.214440} {\bibfield  {journal}
  {\bibinfo  {journal} {Phys. Rev. B}\ }\textbf {\bibinfo {volume} {79}},\
  \bibinfo {pages} {214440} (\bibinfo {year} {2009})}\BibitemShut {NoStop}%
\bibitem [{\citenamefont {Cobanera}\ \emph {et~al.}(2010)\citenamefont
  {Cobanera}, \citenamefont {Ortiz},\ and\ \citenamefont
  {Nussinov}}]{Cobanera_Ortiz_Nussinov_2010}%
  \BibitemOpen
  \bibfield  {author} {\bibinfo {author} {\bibfnamefont {E.}~\bibnamefont
  {Cobanera}}, \bibinfo {author} {\bibfnamefont {G.}~\bibnamefont {Ortiz}},\
  and\ \bibinfo {author} {\bibfnamefont {Z.}~\bibnamefont {Nussinov}},\
  }\bibfield  {title} {\bibinfo {title} {Unified approach to quantum and
  classical dualities},\ }\href
  {https://doi.org/10.1103/PhysRevLett.104.020402} {\bibfield  {journal}
  {\bibinfo  {journal} {Phys. Rev. Lett.}\ }\textbf {\bibinfo {volume} {104}},\
  \bibinfo {pages} {020402} (\bibinfo {year} {2010})}\BibitemShut {NoStop}%
\bibitem [{\citenamefont {Cobanera}\ \emph {et~al.}(2011)\citenamefont
  {Cobanera}, \citenamefont {Ortiz},\ and\ \citenamefont
  {Nussinov}}]{Cobanera_Ortiz_Nussinov_2011}%
  \BibitemOpen
  \bibfield  {author} {\bibinfo {author} {\bibfnamefont {E.}~\bibnamefont
  {Cobanera}}, \bibinfo {author} {\bibfnamefont {G.}~\bibnamefont {Ortiz}},\
  and\ \bibinfo {author} {\bibfnamefont {Z.}~\bibnamefont {Nussinov}},\
  }\bibfield  {title} {\bibinfo {title} {The bond-algebraic approach to
  dualities},\ }\href {https://doi.org/10.1080/00018732.2011.619814} {\bibfield
   {journal} {\bibinfo  {journal} {Advances in Physics}\ }\textbf {\bibinfo
  {volume} {60}},\ \bibinfo {pages} {679–798} (\bibinfo {year} {2011})},\
  \bibinfo {note} {see also the references therein}\BibitemShut {NoStop}%
\bibitem [{\citenamefont {Fendley}(2016)}]{Fendley_2016}%
  \BibitemOpen
  \bibfield  {author} {\bibinfo {author} {\bibfnamefont {P.}~\bibnamefont
  {Fendley}},\ }\bibfield  {title} {\bibinfo {title} {Strong zero modes and
  eigenstate phase transitions in the xyz/interacting majorana chain},\ }\href
  {https://doi.org/10.1088/1751-8113/49/30/30lt01} {\bibfield  {journal}
  {\bibinfo  {journal} {Journal of Physics A: Mathematical and Theoretical}\
  }\textbf {\bibinfo {volume} {49}},\ \bibinfo {pages} {30LT01} (\bibinfo
  {year} {2016})}\BibitemShut {NoStop}%
\bibitem [{\citenamefont {Nussinov}\ and\ \citenamefont
  {Ortiz}(2023)}]{Nussinov_Ortiz_prb2023}%
  \BibitemOpen
  \bibfield  {author} {\bibinfo {author} {\bibfnamefont {Z.}~\bibnamefont
  {Nussinov}}\ and\ \bibinfo {author} {\bibfnamefont {G.}~\bibnamefont
  {Ortiz}},\ }\bibfield  {title} {\bibinfo {title} {Theorem on extensive
  spectral degeneracy for systems with rigid higher symmetries in general
  dimensions},\ }\href {https://doi.org/10.1103/PhysRevB.107.045109} {\bibfield
   {journal} {\bibinfo  {journal} {Phys. Rev. B}\ }\textbf {\bibinfo {volume}
  {107}},\ \bibinfo {pages} {045109} (\bibinfo {year} {2023})},\ \bibinfo
  {note} {and references therein}\BibitemShut {NoStop}%
\bibitem [{\citenamefont {Shvanskaya}\ and\ \citenamefont
  {Vasiliev}(2024)}]{Shvanskaya_Vasiliev_2024}%
  \BibitemOpen
  \bibfield  {author} {\bibinfo {author} {\bibfnamefont {L.~V.}\ \bibnamefont
  {Shvanskaya}}\ and\ \bibinfo {author} {\bibfnamefont {A.~N.}\ \bibnamefont
  {Vasiliev}},\ }\bibfield  {title} {\bibinfo {title} {Diverse magnetic chains
  in inorganic compounds},\ }\href {https://doi.org/10.1021/accountsmr.4c00083}
  {\bibfield  {journal} {\bibinfo  {journal} {Accounts of Materials Research}\
  }\textbf {\bibinfo {volume} {5}},\ \bibinfo {pages} {836–845} (\bibinfo
  {year} {2024})}\BibitemShut {NoStop}%
\bibitem [{\citenamefont {Chepiga}\ and\ \citenamefont
  {White}(2020)}]{Chepiga_White_2020}%
  \BibitemOpen
  \bibfield  {author} {\bibinfo {author} {\bibfnamefont {N.}~\bibnamefont
  {Chepiga}}\ and\ \bibinfo {author} {\bibfnamefont {S.}~\bibnamefont
  {White}},\ }\bibfield  {title} {\bibinfo {title} {Critical properties of a
  comb lattice},\ }\bibfield  {journal} {\bibinfo  {journal} {SciPost Physics}\
  }\textbf {\bibinfo {volume} {9}},\ \href
  {https://doi.org/10.21468/scipostphys.9.1.013} {10.21468/scipostphys.9.1.013}
  (\bibinfo {year} {2020})\BibitemShut {NoStop}%
\bibitem [{\citenamefont {Bhattacharya}\ \emph {et~al.}(2021)\citenamefont
  {Bhattacharya}, \citenamefont {Buser}, \citenamefont {Chandrasekharan},
  \citenamefont {Gupta},\ and\ \citenamefont {Singh}}]{Bhattacharya_etal_2021}%
  \BibitemOpen
  \bibfield  {author} {\bibinfo {author} {\bibfnamefont {T.}~\bibnamefont
  {Bhattacharya}}, \bibinfo {author} {\bibfnamefont {A.~J.}\ \bibnamefont
  {Buser}}, \bibinfo {author} {\bibfnamefont {S.}~\bibnamefont
  {Chandrasekharan}}, \bibinfo {author} {\bibfnamefont {R.}~\bibnamefont
  {Gupta}},\ and\ \bibinfo {author} {\bibfnamefont {H.}~\bibnamefont {Singh}},\
  }\bibfield  {title} {\bibinfo {title} {Qubit regularization of asymptotic
  freedom},\ }\href {https://doi.org/10.1103/PhysRevLett.126.172001} {\bibfield
   {journal} {\bibinfo  {journal} {Phys. Rev. Lett.}\ }\textbf {\bibinfo
  {volume} {126}},\ \bibinfo {pages} {172001} (\bibinfo {year}
  {2021})}\BibitemShut {NoStop}%
\bibitem [{\citenamefont
  {Chandrasekharan}(2025)}]{Chandrasekharan_plenary_2025}%
  \BibitemOpen
  \bibfield  {author} {\bibinfo {author} {\bibfnamefont {S.}~\bibnamefont
  {Chandrasekharan}},\ }\href {https://doi.org/10.48550/ARXIV.2502.05716}
  {\bibinfo {title} {Qubit regularization of quantum field theories}} (\bibinfo
  {year} {2025})\BibitemShut {NoStop}%
\bibitem [{\citenamefont {Jackeli}\ and\ \citenamefont
  {Khaliullin}(2009)}]{Jackeli_Khaliullin_PRL_2009}%
  \BibitemOpen
  \bibfield  {author} {\bibinfo {author} {\bibfnamefont {G.}~\bibnamefont
  {Jackeli}}\ and\ \bibinfo {author} {\bibfnamefont {G.}~\bibnamefont
  {Khaliullin}},\ }\bibfield  {title} {\bibinfo {title} {Mott insulators in the
  strong spin-orbit coupling limit: From heisenberg to a quantum compass and
  kitaev models},\ }\href {https://doi.org/10.1103/PhysRevLett.102.017205}
  {\bibfield  {journal} {\bibinfo  {journal} {Phys. Rev. Lett.}\ }\textbf
  {\bibinfo {volume} {102}},\ \bibinfo {pages} {017205} (\bibinfo {year}
  {2009})}\BibitemShut {NoStop}%
\bibitem [{\citenamefont {Brzezicki}\ and\ \citenamefont
  {Oleś}(2009)}]{Brzezicki_Oles_2009}%
  \BibitemOpen
  \bibfield  {author} {\bibinfo {author} {\bibfnamefont {W.}~\bibnamefont
  {Brzezicki}}\ and\ \bibinfo {author} {\bibfnamefont {A.~M.}\ \bibnamefont
  {Oleś}},\ }\bibfield  {title} {\bibinfo {title} {Exact solution for a
  quantum compass ladder},\ }\bibfield  {journal} {\bibinfo  {journal}
  {Physical Review B}\ }\textbf {\bibinfo {volume} {80}},\ \href
  {https://doi.org/10.1103/physrevb.80.014405} {10.1103/physrevb.80.014405}
  (\bibinfo {year} {2009})\BibitemShut {NoStop}%
\bibitem [{\citenamefont {Patil}\ and\ \citenamefont
  {Sandvik}(2020)}]{Patil_Sandvik_2020}%
  \BibitemOpen
  \bibfield  {author} {\bibinfo {author} {\bibfnamefont {P.}~\bibnamefont
  {Patil}}\ and\ \bibinfo {author} {\bibfnamefont {A.~W.}\ \bibnamefont
  {Sandvik}},\ }\bibfield  {title} {\bibinfo {title} {Hilbert space
  fragmentation and ashkin-teller criticality in fluctuation coupled ising
  models},\ }\href {https://doi.org/10.1103/PhysRevB.101.014453} {\bibfield
  {journal} {\bibinfo  {journal} {Phys. Rev. B}\ }\textbf {\bibinfo {volume}
  {101}},\ \bibinfo {pages} {014453} (\bibinfo {year} {2020})}\BibitemShut
  {NoStop}%
\bibitem [{\citenamefont {Mikeska}\ and\ \citenamefont
  {Kolezhuk}(2004)}]{1d_magnetism_review_2004}%
  \BibitemOpen
  \bibfield  {author} {\bibinfo {author} {\bibfnamefont {H.-J.}\ \bibnamefont
  {Mikeska}}\ and\ \bibinfo {author} {\bibfnamefont {A.~K.}\ \bibnamefont
  {Kolezhuk}},\ }\bibinfo {title} {One-dimensional magnetism},\ in\ \href
  {https://doi.org/10.1007/bfb0119591} {\emph {\bibinfo {booktitle} {Quantum
  Magnetism}}}\ (\bibinfo  {publisher} {Springer Berlin Heidelberg},\ \bibinfo
  {year} {2004})\ p.\ \bibinfo {pages} {1–83}\BibitemShut {NoStop}%
\bibitem [{\citenamefont {Vasiliev}\ \emph {et~al.}(2018)\citenamefont
  {Vasiliev}, \citenamefont {Volkova}, \citenamefont {Zvereva},\ and\
  \citenamefont {Markina}}]{1d_magnetism_review_2018}%
  \BibitemOpen
  \bibfield  {author} {\bibinfo {author} {\bibfnamefont {A.}~\bibnamefont
  {Vasiliev}}, \bibinfo {author} {\bibfnamefont {O.}~\bibnamefont {Volkova}},
  \bibinfo {author} {\bibfnamefont {E.}~\bibnamefont {Zvereva}},\ and\ \bibinfo
  {author} {\bibfnamefont {M.}~\bibnamefont {Markina}},\ }\bibfield  {title}
  {\bibinfo {title} {Milestones of low-d quantum magnetism},\ }\bibfield
  {journal} {\bibinfo  {journal} {npj Quantum Materials}\ }\textbf {\bibinfo
  {volume} {3}},\ \href {https://doi.org/10.1038/s41535-018-0090-7}
  {10.1038/s41535-018-0090-7} (\bibinfo {year} {2018})\BibitemShut {NoStop}%
\bibitem [{\citenamefont {Kitaev}(2003)}]{Kitaev_2003}%
  \BibitemOpen
  \bibfield  {author} {\bibinfo {author} {\bibfnamefont {A.}~\bibnamefont
  {Kitaev}},\ }\bibfield  {title} {\bibinfo {title} {Fault-tolerant quantum
  computation by anyons},\ }\href
  {https://doi.org/10.1016/s0003-4916(02)00018-0} {\bibfield  {journal}
  {\bibinfo  {journal} {Annals of Physics}\ }\textbf {\bibinfo {volume}
  {303}},\ \bibinfo {pages} {2–30} (\bibinfo {year} {2003})}\BibitemShut
  {NoStop}%
\bibitem [{\citenamefont {Kitaev}(2006)}]{Kitaev_2006}%
  \BibitemOpen
  \bibfield  {author} {\bibinfo {author} {\bibfnamefont {A.}~\bibnamefont
  {Kitaev}},\ }\bibfield  {title} {\bibinfo {title} {Anyons in an exactly
  solved model and beyond},\ }\href
  {https://doi.org/https://doi.org/10.1016/j.aop.2005.10.005} {\bibfield
  {journal} {\bibinfo  {journal} {Annals of Physics}\ }\textbf {\bibinfo
  {volume} {321}},\ \bibinfo {pages} {2} (\bibinfo {year} {2006})},\ \bibinfo
  {note} {january Special Issue}\BibitemShut {NoStop}%
\bibitem [{\citenamefont {Khomskii}\ and\ \citenamefont
  {Kugel}(1973)}]{Kugel_Khomskii_1973}%
  \BibitemOpen
  \bibfield  {author} {\bibinfo {author} {\bibfnamefont {D.}~\bibnamefont
  {Khomskii}}\ and\ \bibinfo {author} {\bibfnamefont {K.}~\bibnamefont
  {Kugel}},\ }\bibfield  {title} {\bibinfo {title} {Orbital and magnetic
  structure of two-dimensional ferromagnets with jahn-teller ions},\ }\href
  {https://doi.org/10.1016/0038-1098(73)90362-1} {\bibfield  {journal}
  {\bibinfo  {journal} {Solid State Communications}\ }\textbf {\bibinfo
  {volume} {13}},\ \bibinfo {pages} {763–766} (\bibinfo {year}
  {1973})}\BibitemShut {NoStop}%
\bibitem [{\citenamefont {Kugel’}\ and\ \citenamefont
  {Khomskiĭ}(1982)}]{Kugel_Khomskii_1982}%
  \BibitemOpen
  \bibfield  {author} {\bibinfo {author} {\bibfnamefont {K.~I.}\ \bibnamefont
  {Kugel’}}\ and\ \bibinfo {author} {\bibfnamefont {D.~I.}\ \bibnamefont
  {Khomskiĭ}},\ }\bibfield  {title} {\bibinfo {title} {The jahn-teller effect
  and magnetism: transition metal compounds},\ }\href
  {https://doi.org/10.1070/pu1982v025n04abeh004537} {\bibfield  {journal}
  {\bibinfo  {journal} {Soviet Physics Uspekhi}\ }\textbf {\bibinfo {volume}
  {25}},\ \bibinfo {pages} {231–256} (\bibinfo {year} {1982})}\BibitemShut
  {NoStop}%
\bibitem [{\citenamefont {Dou\c{c}ot}\ \emph {et~al.}(2005)\citenamefont
  {Dou\c{c}ot}, \citenamefont {Feigel'man}, \citenamefont {Ioffe},\ and\
  \citenamefont {Ioselevich}}]{Doucot_etal_2005}%
  \BibitemOpen
  \bibfield  {author} {\bibinfo {author} {\bibfnamefont {B.}~\bibnamefont
  {Dou\c{c}ot}}, \bibinfo {author} {\bibfnamefont {M.~V.}\ \bibnamefont
  {Feigel'man}}, \bibinfo {author} {\bibfnamefont {L.~B.}\ \bibnamefont
  {Ioffe}},\ and\ \bibinfo {author} {\bibfnamefont {A.~S.}\ \bibnamefont
  {Ioselevich}},\ }\bibfield  {title} {\bibinfo {title} {Protected qubits and
  chern-simons theories in josephson junction arrays},\ }\href
  {https://doi.org/10.1103/PhysRevB.71.024505} {\bibfield  {journal} {\bibinfo
  {journal} {Phys. Rev. B}\ }\textbf {\bibinfo {volume} {71}},\ \bibinfo
  {pages} {024505} (\bibinfo {year} {2005})}\BibitemShut {NoStop}%
\bibitem [{\citenamefont {Dorier}\ \emph {et~al.}(2005)\citenamefont {Dorier},
  \citenamefont {Becca},\ and\ \citenamefont {Mila}}]{Dorier_Becca_Mila_2005}%
  \BibitemOpen
  \bibfield  {author} {\bibinfo {author} {\bibfnamefont {J.}~\bibnamefont
  {Dorier}}, \bibinfo {author} {\bibfnamefont {F.}~\bibnamefont {Becca}},\ and\
  \bibinfo {author} {\bibfnamefont {F.}~\bibnamefont {Mila}},\ }\bibfield
  {title} {\bibinfo {title} {Quantum compass model on the square lattice},\
  }\href {https://doi.org/10.1103/PhysRevB.72.024448} {\bibfield  {journal}
  {\bibinfo  {journal} {Phys. Rev. B}\ }\textbf {\bibinfo {volume} {72}},\
  \bibinfo {pages} {024448} (\bibinfo {year} {2005})}\BibitemShut {NoStop}%
\bibitem [{\citenamefont {Nussinov}\ and\ \citenamefont {van~den
  Brink}(2015)}]{Nussinov_vandenBrink_review_2015}%
  \BibitemOpen
  \bibfield  {author} {\bibinfo {author} {\bibfnamefont {Z.}~\bibnamefont
  {Nussinov}}\ and\ \bibinfo {author} {\bibfnamefont {J.}~\bibnamefont {van~den
  Brink}},\ }\bibfield  {title} {\bibinfo {title} {Compass models: Theory and
  physical motivations},\ }\href {https://doi.org/10.1103/RevModPhys.87.1}
  {\bibfield  {journal} {\bibinfo  {journal} {Rev. Mod. Phys.}\ }\textbf
  {\bibinfo {volume} {87}},\ \bibinfo {pages} {1} (\bibinfo {year}
  {2015})}\BibitemShut {NoStop}%
\bibitem [{\citenamefont {Brzezicki}\ \emph {et~al.}(2007)\citenamefont
  {Brzezicki}, \citenamefont {Dziarmaga},\ and\ \citenamefont
  {Ole\'{s}}}]{Brzezicki_Dziarmaga_Oles_2007}%
  \BibitemOpen
  \bibfield  {author} {\bibinfo {author} {\bibfnamefont {W.}~\bibnamefont
  {Brzezicki}}, \bibinfo {author} {\bibfnamefont {J.}~\bibnamefont
  {Dziarmaga}},\ and\ \bibinfo {author} {\bibfnamefont {A.~M.}\ \bibnamefont
  {Ole\'{s}}},\ }\bibfield  {title} {\bibinfo {title} {Quantum phase transition
  in the one-dimensional compass model},\ }\href
  {https://doi.org/10.1103/PhysRevB.75.134415} {\bibfield  {journal} {\bibinfo
  {journal} {Phys. Rev. B}\ }\textbf {\bibinfo {volume} {75}},\ \bibinfo
  {pages} {134415} (\bibinfo {year} {2007})}\BibitemShut {NoStop}%
\bibitem [{foo({\natexlab{a}})}]{footnote_BDO07}%
  \BibitemOpen
  \href@noop {} {} ({\natexlab{a}}),\ \bibinfo {note}
  {ref.~\cite{Brzezicki_Dziarmaga_Oles_2007} had pointed out this degeneracy
  only for the thermodynamic limit using different arguments.}\BibitemShut
  {Stop}%
\bibitem [{\citenamefont {Sch\"afer}\ \emph {et~al.}(2023)\citenamefont
  {Sch\"afer}, \citenamefont {Placke}, \citenamefont {Benton},\ and\
  \citenamefont {Moessner}}]{Schafer_etal_2023}%
  \BibitemOpen
  \bibfield  {author} {\bibinfo {author} {\bibfnamefont {R.}~\bibnamefont
  {Sch\"afer}}, \bibinfo {author} {\bibfnamefont {B.}~\bibnamefont {Placke}},
  \bibinfo {author} {\bibfnamefont {O.}~\bibnamefont {Benton}},\ and\ \bibinfo
  {author} {\bibfnamefont {R.}~\bibnamefont {Moessner}},\ }\bibfield  {title}
  {\bibinfo {title} {Abundance of hard-hexagon crystals in the quantum
  pyrochlore antiferromagnet},\ }\href
  {https://doi.org/10.1103/PhysRevLett.131.096702} {\bibfield  {journal}
  {\bibinfo  {journal} {Phys. Rev. Lett.}\ }\textbf {\bibinfo {volume} {131}},\
  \bibinfo {pages} {096702} (\bibinfo {year} {2023})}\BibitemShut {NoStop}%
\bibitem [{\citenamefont {Baskaran}\ \emph {et~al.}(2007)\citenamefont
  {Baskaran}, \citenamefont {Mandal},\ and\ \citenamefont
  {Shankar}}]{Baskaran_Mandal_Shankar_PRL_2007}%
  \BibitemOpen
  \bibfield  {author} {\bibinfo {author} {\bibfnamefont {G.}~\bibnamefont
  {Baskaran}}, \bibinfo {author} {\bibfnamefont {S.}~\bibnamefont {Mandal}},\
  and\ \bibinfo {author} {\bibfnamefont {R.}~\bibnamefont {Shankar}},\
  }\bibfield  {title} {\bibinfo {title} {Exact results for spin dynamics and
  fractionalization in the kitaev model},\ }\href
  {https://doi.org/10.1103/PhysRevLett.98.247201} {\bibfield  {journal}
  {\bibinfo  {journal} {Phys. Rev. Lett.}\ }\textbf {\bibinfo {volume} {98}},\
  \bibinfo {pages} {247201} (\bibinfo {year} {2007})}\BibitemShut {NoStop}%
\bibitem [{\citenamefont {Chen}\ and\ \citenamefont
  {Nussinov}(2008)}]{Nussinov_Chen_2008}%
  \BibitemOpen
  \bibfield  {author} {\bibinfo {author} {\bibfnamefont {H.-D.}\ \bibnamefont
  {Chen}}\ and\ \bibinfo {author} {\bibfnamefont {Z.}~\bibnamefont
  {Nussinov}},\ }\bibfield  {title} {\bibinfo {title} {Exact results of the
  kitaev model on a hexagonal lattice: spin states, string and brane
  correlators, and anyonic excitations},\ }\href
  {https://doi.org/10.1088/1751-8113/41/7/075001} {\bibfield  {journal}
  {\bibinfo  {journal} {Journal of Physics A: Mathematical and Theoretical}\
  }\textbf {\bibinfo {volume} {41}},\ \bibinfo {pages} {075001} (\bibinfo
  {year} {2008})}\BibitemShut {NoStop}%
\bibitem [{\citenamefont {Haah}(2021)}]{Haah_2021}%
  \BibitemOpen
  \bibfield  {author} {\bibinfo {author} {\bibfnamefont {J.}~\bibnamefont
  {Haah}},\ }\bibfield  {title} {\bibinfo {title} {A degeneracy bound for
  homogeneous topological order},\ }\bibfield  {journal} {\bibinfo  {journal}
  {SciPost Physics}\ }\textbf {\bibinfo {volume} {10}},\ \href
  {https://doi.org/10.21468/scipostphys.10.1.011}
  {10.21468/scipostphys.10.1.011} (\bibinfo {year} {2021})\BibitemShut
  {NoStop}%
\bibitem [{slo()}]{slow_mode}%
  \BibitemOpen
  \bibinfo {note} {See the conjecture on the development of a slow mode in
  presence of additional perturbations in Ref.~\cite{JWpreprint}.}\BibitemShut
  {Stop}%
\bibitem [{Sap()}]{Saptarshi_private}%
  \BibitemOpen
  \bibinfo {note} {Saptarshi Mandal, private discussion, IOP Bhubaneshwar,
  Auguest 2024.}\BibitemShut {Stop}%
\bibitem [{\citenamefont {Mandal}\ \emph {et~al.}(2012)\citenamefont {Mandal},
  \citenamefont {Shankar},\ and\ \citenamefont
  {Baskaran}}]{Mandal_Shankar_Baskaran_2012}%
  \BibitemOpen
  \bibfield  {author} {\bibinfo {author} {\bibfnamefont {S.}~\bibnamefont
  {Mandal}}, \bibinfo {author} {\bibfnamefont {R.}~\bibnamefont {Shankar}},\
  and\ \bibinfo {author} {\bibfnamefont {G.}~\bibnamefont {Baskaran}},\
  }\bibfield  {title} {\bibinfo {title} {Rvb gauge theory and the topological
  degeneracy in the honeycomb kitaev model},\ }\href
  {https://doi.org/10.1088/1751-8113/45/33/335304} {\bibfield  {journal}
  {\bibinfo  {journal} {Journal of Physics A: Mathematical and Theoretical}\
  }\textbf {\bibinfo {volume} {45}},\ \bibinfo {pages} {335304} (\bibinfo
  {year} {2012})}\BibitemShut {NoStop}%
\bibitem [{\citenamefont {Levin}\ and\ \citenamefont
  {Wen}(2005)}]{Levin_Wen_2005}%
  \BibitemOpen
  \bibfield  {author} {\bibinfo {author} {\bibfnamefont {M.~A.}\ \bibnamefont
  {Levin}}\ and\ \bibinfo {author} {\bibfnamefont {X.-G.}\ \bibnamefont
  {Wen}},\ }\bibfield  {title} {\bibinfo {title} {String-net condensation: A
  physical mechanism for topological phases},\ }\href
  {https://doi.org/10.1103/PhysRevB.71.045110} {\bibfield  {journal} {\bibinfo
  {journal} {Phys. Rev. B}\ }\textbf {\bibinfo {volume} {71}},\ \bibinfo
  {pages} {045110} (\bibinfo {year} {2005})}\BibitemShut {NoStop}%
\bibitem [{\citenamefont {Haah}(2011)}]{Haah_2011}%
  \BibitemOpen
  \bibfield  {author} {\bibinfo {author} {\bibfnamefont {J.}~\bibnamefont
  {Haah}},\ }\bibfield  {title} {\bibinfo {title} {Local stabilizer codes in
  three dimensions without string logical operators},\ }\href
  {https://doi.org/10.1103/PhysRevA.83.042330} {\bibfield  {journal} {\bibinfo
  {journal} {Phys. Rev. A}\ }\textbf {\bibinfo {volume} {83}},\ \bibinfo
  {pages} {042330} (\bibinfo {year} {2011})}\BibitemShut {NoStop}%
\bibitem [{\citenamefont {Vijay}\ \emph {et~al.}(2016)\citenamefont {Vijay},
  \citenamefont {Haah},\ and\ \citenamefont {Fu}}]{Vijay_Haah_Fu_2016}%
  \BibitemOpen
  \bibfield  {author} {\bibinfo {author} {\bibfnamefont {S.}~\bibnamefont
  {Vijay}}, \bibinfo {author} {\bibfnamefont {J.}~\bibnamefont {Haah}},\ and\
  \bibinfo {author} {\bibfnamefont {L.}~\bibnamefont {Fu}},\ }\bibfield
  {title} {\bibinfo {title} {Fracton topological order, generalized lattice
  gauge theory, and duality},\ }\href
  {https://doi.org/10.1103/PhysRevB.94.235157} {\bibfield  {journal} {\bibinfo
  {journal} {Phys. Rev. B}\ }\textbf {\bibinfo {volume} {94}},\ \bibinfo
  {pages} {235157} (\bibinfo {year} {2016})}\BibitemShut {NoStop}%
\bibitem [{\citenamefont {Pretko}\ \emph {et~al.}(2020)\citenamefont {Pretko},
  \citenamefont {Chen},\ and\ \citenamefont {You}}]{Pretko_Chen_You_2020}%
  \BibitemOpen
  \bibfield  {author} {\bibinfo {author} {\bibfnamefont {M.}~\bibnamefont
  {Pretko}}, \bibinfo {author} {\bibfnamefont {X.}~\bibnamefont {Chen}},\ and\
  \bibinfo {author} {\bibfnamefont {Y.}~\bibnamefont {You}},\ }\bibfield
  {title} {\bibinfo {title} {Fracton phases of matter},\ }\href
  {https://doi.org/10.1142/S0217751X20300033} {\bibfield  {journal} {\bibinfo
  {journal} {International Journal of Modern Physics A}\ }\textbf {\bibinfo
  {volume} {35}},\ \bibinfo {pages} {2030003} (\bibinfo {year} {2020})},\
  \Eprint {https://arxiv.org/abs/https://doi.org/10.1142/S0217751X20300033}
  {https://doi.org/10.1142/S0217751X20300033} \BibitemShut {NoStop}%
\bibitem [{foo({\natexlab{b}})}]{footnote_SYK}%
  \BibitemOpen
  \bibinfo {note} {The ground state entropy of SYK model appears to have a
  different origin from this perspective, however the general approach of
  analysing through the conserved quantitites and their mutual algebra might
  still be relevant for SYK model.}\BibitemShut {Stop}%
\bibitem [{\citenamefont {Homeier}\ \emph {et~al.}(2021)\citenamefont
  {Homeier}, \citenamefont {Schweizer}, \citenamefont {Aidelsburger},
  \citenamefont {Fedorov},\ and\ \citenamefont {Grusdt}}]{Homeier_etal_2021}%
  \BibitemOpen
  \bibfield  {author} {\bibinfo {author} {\bibfnamefont {L.}~\bibnamefont
  {Homeier}}, \bibinfo {author} {\bibfnamefont {C.}~\bibnamefont {Schweizer}},
  \bibinfo {author} {\bibfnamefont {M.}~\bibnamefont {Aidelsburger}}, \bibinfo
  {author} {\bibfnamefont {A.}~\bibnamefont {Fedorov}},\ and\ \bibinfo {author}
  {\bibfnamefont {F.}~\bibnamefont {Grusdt}},\ }\bibfield  {title} {\bibinfo
  {title} {$\mathbb{Z}_2$ lattice gauge theories and kitaev’s toric code: A
  scheme for analog quantum simulation},\ }\bibfield  {journal} {\bibinfo
  {journal} {Physical Review B}\ }\textbf {\bibinfo {volume} {104}},\ \href
  {https://doi.org/10.1103/physrevb.104.085138} {10.1103/physrevb.104.085138}
  (\bibinfo {year} {2021})\BibitemShut {NoStop}%
\bibitem [{\citenamefont {Shah}\ \emph {et~al.}(2025)\citenamefont {Shah},
  \citenamefont {Nambiar}, \citenamefont {Gorshkov},\ and\ \citenamefont
  {Galitski}}]{Shah_etal_2025}%
  \BibitemOpen
  \bibfield  {author} {\bibinfo {author} {\bibfnamefont {J.}~\bibnamefont
  {Shah}}, \bibinfo {author} {\bibfnamefont {G.}~\bibnamefont {Nambiar}},
  \bibinfo {author} {\bibfnamefont {A.~V.}\ \bibnamefont {Gorshkov}},\ and\
  \bibinfo {author} {\bibfnamefont {V.}~\bibnamefont {Galitski}},\ }\bibfield
  {title} {\bibinfo {title} {Quantum spin ice in three-dimensional rydberg atom
  arrays},\ }\bibfield  {journal} {\bibinfo  {journal} {Physical Review X}\
  }\textbf {\bibinfo {volume} {15}},\ \href
  {https://doi.org/10.1103/physrevx.15.011025} {10.1103/physrevx.15.011025}
  (\bibinfo {year} {2025})\BibitemShut {NoStop}%
\bibitem [{\citenamefont {Agrapidis}\ \emph {et~al.}(2018)\citenamefont
  {Agrapidis}, \citenamefont {van~den Brink},\ and\ \citenamefont
  {Nishimoto}}]{Agrapidis_Brink_Nishimoto_2018}%
  \BibitemOpen
  \bibfield  {author} {\bibinfo {author} {\bibfnamefont {C.~E.}\ \bibnamefont
  {Agrapidis}}, \bibinfo {author} {\bibfnamefont {J.}~\bibnamefont {van~den
  Brink}},\ and\ \bibinfo {author} {\bibfnamefont {S.}~\bibnamefont
  {Nishimoto}},\ }\bibfield  {title} {\bibinfo {title} {Ordered states in the
  kitaev-heisenberg model: From 1d chains to 2d honeycomb},\ }\href@noop {}
  {\bibfield  {journal} {\bibinfo  {journal} {Sci. Rep.}\ }\textbf {\bibinfo
  {volume} {8}},\ \bibinfo {pages} {1815} (\bibinfo {year} {2018})}\BibitemShut
  {NoStop}%
\bibitem [{\citenamefont {Morris}\ \emph {et~al.}(2021)\citenamefont {Morris},
  \citenamefont {Desai}, \citenamefont {Viirok}, \citenamefont {H{\"u}vonen},
  \citenamefont {Nagel}, \citenamefont {R{\~o}{\~o}m}, \citenamefont {Krizan},
  \citenamefont {Cava}, \citenamefont {McQueen}, \citenamefont {Koohpayeh},
  \citenamefont {Kaul},\ and\ \citenamefont {Armitage}}]{Morris_etal_2021}%
  \BibitemOpen
  \bibfield  {author} {\bibinfo {author} {\bibfnamefont {C.~M.}\ \bibnamefont
  {Morris}}, \bibinfo {author} {\bibfnamefont {N.}~\bibnamefont {Desai}},
  \bibinfo {author} {\bibfnamefont {J.}~\bibnamefont {Viirok}}, \bibinfo
  {author} {\bibfnamefont {D.}~\bibnamefont {H{\"u}vonen}}, \bibinfo {author}
  {\bibfnamefont {U.}~\bibnamefont {Nagel}}, \bibinfo {author} {\bibfnamefont
  {T.}~\bibnamefont {R{\~o}{\~o}m}}, \bibinfo {author} {\bibfnamefont {J.~W.}\
  \bibnamefont {Krizan}}, \bibinfo {author} {\bibfnamefont {R.~J.}\
  \bibnamefont {Cava}}, \bibinfo {author} {\bibfnamefont {T.~M.}\ \bibnamefont
  {McQueen}}, \bibinfo {author} {\bibfnamefont {S.~M.}\ \bibnamefont
  {Koohpayeh}}, \bibinfo {author} {\bibfnamefont {R.~K.}\ \bibnamefont
  {Kaul}},\ and\ \bibinfo {author} {\bibfnamefont {N.~P.}\ \bibnamefont
  {Armitage}},\ }\bibfield  {title} {\bibinfo {title} {Duality and domain wall
  dynamics in a twisted kitaev chain},\ }\href@noop {} {\bibfield  {journal}
  {\bibinfo  {journal} {Nat. Phys.}\ }\textbf {\bibinfo {volume} {17}},\
  \bibinfo {pages} {832} (\bibinfo {year} {2021})}\BibitemShut {NoStop}%
\bibitem [{\citenamefont {Fu}\ \emph {et~al.}(2018)\citenamefont {Fu},
  \citenamefont {Knolle},\ and\ \citenamefont
  {Perkins}}]{Fu_Knolle_Perkins_PRB_2018}%
  \BibitemOpen
  \bibfield  {author} {\bibinfo {author} {\bibfnamefont {J.}~\bibnamefont
  {Fu}}, \bibinfo {author} {\bibfnamefont {J.}~\bibnamefont {Knolle}},\ and\
  \bibinfo {author} {\bibfnamefont {N.~B.}\ \bibnamefont {Perkins}},\
  }\bibfield  {title} {\bibinfo {title} {Three types of representation of spin
  in terms of majorana fermions and an alternative solution of the kitaev
  honeycomb model},\ }\href {https://doi.org/10.1103/PhysRevB.97.115142}
  {\bibfield  {journal} {\bibinfo  {journal} {Phys. Rev. B}\ }\textbf {\bibinfo
  {volume} {97}},\ \bibinfo {pages} {115142} (\bibinfo {year}
  {2018})}\BibitemShut {NoStop}%
\bibitem [{\citenamefont {Wenzel}\ and\ \citenamefont
  {Janke}(2009)}]{Wenzel_Janke_2009}%
  \BibitemOpen
  \bibfield  {author} {\bibinfo {author} {\bibfnamefont {S.}~\bibnamefont
  {Wenzel}}\ and\ \bibinfo {author} {\bibfnamefont {W.}~\bibnamefont {Janke}},\
  }\bibfield  {title} {\bibinfo {title} {Finite-temperature n\'eel ordering of
  fluctuations in a plaquette orbital model},\ }\href
  {https://doi.org/10.1103/PhysRevB.80.054403} {\bibfield  {journal} {\bibinfo
  {journal} {Phys. Rev. B}\ }\textbf {\bibinfo {volume} {80}},\ \bibinfo
  {pages} {054403} (\bibinfo {year} {2009})}\BibitemShut {NoStop}%
\bibitem [{\citenamefont {Wenzel}\ and\ \citenamefont
  {Janke}(2008)}]{Wenzel_Janke_2008}%
  \BibitemOpen
  \bibfield  {author} {\bibinfo {author} {\bibfnamefont {S.}~\bibnamefont
  {Wenzel}}\ and\ \bibinfo {author} {\bibfnamefont {W.}~\bibnamefont {Janke}},\
  }\bibfield  {title} {\bibinfo {title} {Monte carlo simulations of the
  directional-ordering transition in the two-dimensional classical and quantum
  compass model},\ }\href {https://doi.org/10.1103/PhysRevB.78.064402}
  {\bibfield  {journal} {\bibinfo  {journal} {Phys. Rev. B}\ }\textbf {\bibinfo
  {volume} {78}},\ \bibinfo {pages} {064402} (\bibinfo {year}
  {2008})}\BibitemShut {NoStop}%
\bibitem [{\citenamefont {Wenzel}\ \emph {et~al.}(2010)\citenamefont {Wenzel},
  \citenamefont {Janke},\ and\ \citenamefont
  {L\"auchli}}]{Wenzel_Janke_Laeuchli_2010}%
  \BibitemOpen
  \bibfield  {author} {\bibinfo {author} {\bibfnamefont {S.}~\bibnamefont
  {Wenzel}}, \bibinfo {author} {\bibfnamefont {W.}~\bibnamefont {Janke}},\ and\
  \bibinfo {author} {\bibfnamefont {A.~M.}\ \bibnamefont {L\"auchli}},\
  }\bibfield  {title} {\bibinfo {title} {Re-examining the directional-ordering
  transition in the compass model with screw-periodic boundary conditions},\
  }\href {https://doi.org/10.1103/PhysRevE.81.066702} {\bibfield  {journal}
  {\bibinfo  {journal} {Phys. Rev. E}\ }\textbf {\bibinfo {volume} {81}},\
  \bibinfo {pages} {066702} (\bibinfo {year} {2010})}\BibitemShut {NoStop}%
\bibitem [{\citenamefont {Biskup}\ and\ \citenamefont
  {Kotecký}(2010)}]{Biskup_Kotecky_2010}%
  \BibitemOpen
  \bibfield  {author} {\bibinfo {author} {\bibfnamefont {M.}~\bibnamefont
  {Biskup}}\ and\ \bibinfo {author} {\bibfnamefont {R.}~\bibnamefont
  {Kotecký}},\ }\bibfield  {title} {\bibinfo {title} {True nature of
  long-range order in a plaquette orbital model},\ }\href
  {https://doi.org/10.1088/1742-5468/2010/11/p11001} {\bibfield  {journal}
  {\bibinfo  {journal} {Journal of Statistical Mechanics: Theory and
  Experiment}\ }\textbf {\bibinfo {volume} {2010}},\ \bibinfo {pages} {P11001}
  (\bibinfo {year} {2010})}\BibitemShut {NoStop}%
\bibitem [{\citenamefont {Ogura}\ \emph {et~al.}(2020)\citenamefont {Ogura},
  \citenamefont {Imamura}, \citenamefont {Kameyama}, \citenamefont {Minami},\
  and\ \citenamefont {Sato}}]{Ogura_etal_2020}%
  \BibitemOpen
  \bibfield  {author} {\bibinfo {author} {\bibfnamefont {M.}~\bibnamefont
  {Ogura}}, \bibinfo {author} {\bibfnamefont {Y.}~\bibnamefont {Imamura}},
  \bibinfo {author} {\bibfnamefont {N.}~\bibnamefont {Kameyama}}, \bibinfo
  {author} {\bibfnamefont {K.}~\bibnamefont {Minami}},\ and\ \bibinfo {author}
  {\bibfnamefont {M.}~\bibnamefont {Sato}},\ }\bibfield  {title} {\bibinfo
  {title} {Geometric criterion for solvability of lattice spin systems},\
  }\href {https://doi.org/10.1103/PhysRevB.102.245118} {\bibfield  {journal}
  {\bibinfo  {journal} {Phys. Rev. B}\ }\textbf {\bibinfo {volume} {102}},\
  \bibinfo {pages} {245118} (\bibinfo {year} {2020})}\BibitemShut {NoStop}%
\bibitem [{\citenamefont {Chapman}\ and\ \citenamefont
  {Flammia}(2020)}]{Chapman_Flammia_2020}%
  \BibitemOpen
  \bibfield  {author} {\bibinfo {author} {\bibfnamefont {A.}~\bibnamefont
  {Chapman}}\ and\ \bibinfo {author} {\bibfnamefont {S.~T.}\ \bibnamefont
  {Flammia}},\ }\bibfield  {title} {\bibinfo {title} {Characterization of
  solvable spin models via graph invariants},\ }\href
  {https://doi.org/10.22331/q-2020-06-04-278} {\bibfield  {journal} {\bibinfo
  {journal} {{Quantum}}\ }\textbf {\bibinfo {volume} {4}},\ \bibinfo {pages}
  {278} (\bibinfo {year} {2020})}\BibitemShut {NoStop}%
\bibitem [{\citenamefont {Minami}(2016)}]{Minami_2016}%
  \BibitemOpen
  \bibfield  {author} {\bibinfo {author} {\bibfnamefont {K.}~\bibnamefont
  {Minami}},\ }\bibfield  {title} {\bibinfo {title} {Solvable hamiltonians and
  fermionization transformations obtained from operators satisfying specific
  commutation relations},\ }\href {https://doi.org/10.7566/JPSJ.85.024003}
  {\bibfield  {journal} {\bibinfo  {journal} {Journal of the Physical Society
  of Japan}\ }\textbf {\bibinfo {volume} {85}},\ \bibinfo {pages} {024003}
  (\bibinfo {year} {2016})},\ \Eprint
  {https://arxiv.org/abs/https://doi.org/10.7566/JPSJ.85.024003}
  {https://doi.org/10.7566/JPSJ.85.024003} \BibitemShut {NoStop}%
\bibitem [{\citenamefont {Minami}(2017)}]{Minami_2017}%
  \BibitemOpen
  \bibfield  {author} {\bibinfo {author} {\bibfnamefont {K.}~\bibnamefont
  {Minami}},\ }\bibfield  {title} {\bibinfo {title} {Infinite number of
  solvable generalizations of xy-chain, with cluster state, and with central
  charge c=m/2},\ }\href
  {https://doi.org/https://doi.org/10.1016/j.nuclphysb.2017.10.004} {\bibfield
  {journal} {\bibinfo  {journal} {Nuclear Physics B}\ }\textbf {\bibinfo
  {volume} {925}},\ \bibinfo {pages} {144} (\bibinfo {year}
  {2017})}\BibitemShut {NoStop}%
\bibitem [{\citenamefont {Yanagihara}\ and\ \citenamefont
  {Minami}(2020)}]{Minami_Yanagihara_2020}%
  \BibitemOpen
  \bibfield  {author} {\bibinfo {author} {\bibfnamefont {Y.}~\bibnamefont
  {Yanagihara}}\ and\ \bibinfo {author} {\bibfnamefont {K.}~\bibnamefont
  {Minami}},\ }\bibfield  {title} {\bibinfo {title} {{Exact solution of a
  cluster model with next-nearest-neighbor interaction}},\ }\href
  {https://doi.org/10.1093/ptep/ptaa146} {\bibfield  {journal} {\bibinfo
  {journal} {Progress of Theoretical and Experimental Physics}\ }\textbf
  {\bibinfo {volume} {2020}},\ \bibinfo {pages} {113A01} (\bibinfo {year}
  {2020})},\ \Eprint
  {https://arxiv.org/abs/https://academic.oup.com/ptep/article-pdf/2020/11/113A01/34585154/ptaa146.pdf}
  {https://academic.oup.com/ptep/article-pdf/2020/11/113A01/34585154/ptaa146.pdf}
  \BibitemShut {NoStop}%
\bibitem [{\citenamefont {Elman}\ \emph {et~al.}(2021)\citenamefont {Elman},
  \citenamefont {Chapman},\ and\ \citenamefont
  {Flammia}}]{Chapman_Elman_Flammia_2021}%
  \BibitemOpen
  \bibfield  {author} {\bibinfo {author} {\bibfnamefont {S.~J.}\ \bibnamefont
  {Elman}}, \bibinfo {author} {\bibfnamefont {A.}~\bibnamefont {Chapman}},\
  and\ \bibinfo {author} {\bibfnamefont {S.~T.}\ \bibnamefont {Flammia}},\
  }\bibfield  {title} {\bibinfo {title} {Free fermions behind the disguise},\
  }\href {https://doi.org/10.1007/s00220-021-04220-w} {\bibfield  {journal}
  {\bibinfo  {journal} {Communications in Mathematical Physics}\ }\textbf
  {\bibinfo {volume} {388}},\ \bibinfo {pages} {969} (\bibinfo {year}
  {2021})}\BibitemShut {NoStop}%
\bibitem [{\citenamefont {Chapman}\ \emph {et~al.}(2023)\citenamefont
  {Chapman}, \citenamefont {Elman},\ and\ \citenamefont
  {Mann}}]{Chapman_Elman_Mann_2023}%
  \BibitemOpen
  \bibfield  {author} {\bibinfo {author} {\bibfnamefont {A.}~\bibnamefont
  {Chapman}}, \bibinfo {author} {\bibfnamefont {S.~J.}\ \bibnamefont {Elman}},\
  and\ \bibinfo {author} {\bibfnamefont {R.~L.}\ \bibnamefont {Mann}},\ }\href
  {https://arxiv.org/abs/2305.15625} {\bibinfo {title} {A unified
  graph-theoretic framework for free-fermion solvability}} (\bibinfo {year}
  {2023}),\ \Eprint {https://arxiv.org/abs/2305.15625} {arXiv:2305.15625
  [quant-ph]} \BibitemShut {NoStop}%
\bibitem [{\citenamefont {Fendley}\ and\ \citenamefont
  {Pozsgay}(2024)}]{Fendley_Pozsgay_2024}%
  \BibitemOpen
  \bibfield  {author} {\bibinfo {author} {\bibfnamefont {P.}~\bibnamefont
  {Fendley}}\ and\ \bibinfo {author} {\bibfnamefont {B.}~\bibnamefont
  {Pozsgay}},\ }\bibfield  {title} {\bibinfo {title} {{Free fermions beyond
  Jordan and Wigner}},\ }\href {https://doi.org/10.21468/SciPostPhys.16.4.102}
  {\bibfield  {journal} {\bibinfo  {journal} {SciPost Phys.}\ }\textbf
  {\bibinfo {volume} {16}},\ \bibinfo {pages} {102} (\bibinfo {year}
  {2024})}\BibitemShut {NoStop}%
\bibitem [{\citenamefont {Fendley}(2019)}]{Fendley_2019}%
  \BibitemOpen
  \bibfield  {author} {\bibinfo {author} {\bibfnamefont {P.}~\bibnamefont
  {Fendley}},\ }\bibfield  {title} {\bibinfo {title} {Free fermions in
  disguise},\ }\href {https://doi.org/10.1088/1751-8121/ab305d} {\bibfield
  {journal} {\bibinfo  {journal} {Journal of Physics A: Mathematical and
  Theoretical}\ }\textbf {\bibinfo {volume} {52}},\ \bibinfo {pages} {335002}
  (\bibinfo {year} {2019})}\BibitemShut {NoStop}%
\bibitem [{\citenamefont {Chen}\ \emph {et~al.}(2024)\citenamefont {Chen},
  \citenamefont {Ellison}, \citenamefont {Cheng}, \citenamefont {Ye},\ and\
  \citenamefont {Chen}}]{Chen_etal_2024}%
  \BibitemOpen
  \bibfield  {author} {\bibinfo {author} {\bibfnamefont {L.-M.}\ \bibnamefont
  {Chen}}, \bibinfo {author} {\bibfnamefont {T.~D.}\ \bibnamefont {Ellison}},
  \bibinfo {author} {\bibfnamefont {M.}~\bibnamefont {Cheng}}, \bibinfo
  {author} {\bibfnamefont {P.}~\bibnamefont {Ye}},\ and\ \bibinfo {author}
  {\bibfnamefont {J.-Y.}\ \bibnamefont {Chen}},\ }\bibfield  {title} {\bibinfo
  {title} {Chiral spin liquid in a ${\mathbb{z}}_{3}$ kitaev model},\ }\href
  {https://doi.org/10.1103/PhysRevB.109.155161} {\bibfield  {journal} {\bibinfo
   {journal} {Phys. Rev. B}\ }\textbf {\bibinfo {volume} {109}},\ \bibinfo
  {pages} {155161} (\bibinfo {year} {2024})}\BibitemShut {NoStop}%
\bibitem [{hig()}]{higher_form_footnote}%
  \BibitemOpen
  \bibinfo {note} {See Refs.~[20-23] in
  Ref.~\cite{Chen_etal_2024}.}\BibitemShut {Stop}%
\bibitem [{\citenamefont {Tong}(2016)}]{Tong_notes_2016}%
  \BibitemOpen
  \bibfield  {author} {\bibinfo {author} {\bibfnamefont {D.}~\bibnamefont
  {Tong}},\ }\href {https://doi.org/10.48550/ARXIV.1606.06687} {\bibinfo
  {title} {Lectures on the quantum hall effect}} (\bibinfo {year}
  {2016})\BibitemShut {NoStop}%
\bibitem [{\citenamefont {Maldacena}\ \emph {et~al.}(2016)\citenamefont
  {Maldacena}, \citenamefont {Shenker},\ and\ \citenamefont
  {Stanford}}]{Maldacena_Shenker_Stanford_2016}%
  \BibitemOpen
  \bibfield  {author} {\bibinfo {author} {\bibfnamefont {J.}~\bibnamefont
  {Maldacena}}, \bibinfo {author} {\bibfnamefont {S.~H.}\ \bibnamefont
  {Shenker}},\ and\ \bibinfo {author} {\bibfnamefont {D.}~\bibnamefont
  {Stanford}},\ }\bibfield  {title} {\bibinfo {title} {A bound on chaos},\
  }\href {https://doi.org/10.1007/JHEP08(2016)106} {\bibfield  {journal}
  {\bibinfo  {journal} {Journal of High Energy Physics}\ }\textbf {\bibinfo
  {volume} {2016}},\ \bibinfo {pages} {106} (\bibinfo {year}
  {2016})}\BibitemShut {NoStop}%
\bibitem [{Kit()}]{Kitaev_talk_2015}%
  \BibitemOpen
  \bibinfo {note} {Kitaev, Alexei (2015), “A simple model of quantum
  holography, talk given at kitp program: entanglement in stronglycorrelated
  quantum matter,” USA April 2015.}\BibitemShut {Stop}%
\bibitem [{\citenamefont {Kitaev}\ and\ \citenamefont
  {Suh}(2018)}]{Kitaev_Suh_2018}%
  \BibitemOpen
  \bibfield  {author} {\bibinfo {author} {\bibfnamefont {A.}~\bibnamefont
  {Kitaev}}\ and\ \bibinfo {author} {\bibfnamefont {S.~J.}\ \bibnamefont
  {Suh}},\ }\bibfield  {title} {\bibinfo {title} {The soft mode in the
  sachdev-ye-kitaev model and its gravity dual},\ }\href
  {https://doi.org/10.1007/JHEP05(2018)183} {\bibfield  {journal} {\bibinfo
  {journal} {Journal of High Energy Physics}\ }\textbf {\bibinfo {volume}
  {2018}},\ \bibinfo {pages} {183} (\bibinfo {year} {2018})}\BibitemShut
  {NoStop}%
\bibitem [{\citenamefont {Maldacena}\ and\ \citenamefont
  {Stanford}(2016)}]{Maldacena_Stanford_2016}%
  \BibitemOpen
  \bibfield  {author} {\bibinfo {author} {\bibfnamefont {J.}~\bibnamefont
  {Maldacena}}\ and\ \bibinfo {author} {\bibfnamefont {D.}~\bibnamefont
  {Stanford}},\ }\bibfield  {title} {\bibinfo {title} {Remarks on the
  sachdev-ye-kitaev model},\ }\href
  {https://doi.org/10.1103/PhysRevD.94.106002} {\bibfield  {journal} {\bibinfo
  {journal} {Phys. Rev. D}\ }\textbf {\bibinfo {volume} {94}},\ \bibinfo
  {pages} {106002} (\bibinfo {year} {2016})}\BibitemShut {NoStop}%
\bibitem [{foo({\natexlab{c}})}]{footnote_SYK_review_secVD}%
  \BibitemOpen
  \bibinfo {note} {See Sec. 5D of Ref.~\cite{SYK_review}.}\BibitemShut {Stop}%
\bibitem [{foo({\natexlab{d}})}]{footnote_SYK_review_secXIIE}%
  \BibitemOpen
  \bibinfo {note} {See Sec. 12E of Ref.~\cite{SYK_review}.}\BibitemShut {Stop}%
\bibitem [{\citenamefont {Sachdev}\ and\ \citenamefont
  {Ye}(1993)}]{Sachdev_Ye_1993}%
  \BibitemOpen
  \bibfield  {author} {\bibinfo {author} {\bibfnamefont {S.}~\bibnamefont
  {Sachdev}}\ and\ \bibinfo {author} {\bibfnamefont {J.}~\bibnamefont {Ye}},\
  }\bibfield  {title} {\bibinfo {title} {Gapless spin-fluid ground state in a
  random quantum heisenberg magnet},\ }\href
  {https://doi.org/10.1103/PhysRevLett.70.3339} {\bibfield  {journal} {\bibinfo
   {journal} {Phys. Rev. Lett.}\ }\textbf {\bibinfo {volume} {70}},\ \bibinfo
  {pages} {3339} (\bibinfo {year} {1993})}\BibitemShut {NoStop}%
\bibitem [{\citenamefont {Garcia-Mata}\ \emph {et~al.}(2023)\citenamefont
  {Garcia-Mata}, \citenamefont {Jalabert},\ and\ \citenamefont
  {Wisniacki}}]{OTOC_encyclopedia}%
  \BibitemOpen
  \bibfield  {author} {\bibinfo {author} {\bibfnamefont {I.}~\bibnamefont
  {Garcia-Mata}}, \bibinfo {author} {\bibfnamefont {R.}~\bibnamefont
  {Jalabert}},\ and\ \bibinfo {author} {\bibfnamefont {D.}~\bibnamefont
  {Wisniacki}},\ }\bibfield  {title} {\bibinfo {title} {Out-of-time-order
  correlations and quantum chaos},\ }\href
  {https://doi.org/10.4249/scholarpedia.55237} {\bibfield  {journal} {\bibinfo
  {journal} {Scholarpedia}\ }\textbf {\bibinfo {volume} {18}},\ \bibinfo
  {pages} {55237} (\bibinfo {year} {2023})}\BibitemShut {NoStop}%
\bibitem [{bla()}]{black_hole}%
  \BibitemOpen
  \bibinfo {note} {\textcolor{black}{Another speculation would then be if these
  models with spin-$\frac{1}{2}$ microscopic degrees of freedom or qubits
  connect somehow to black hole physics in analogy with the connection between
  the SYK model and charged black holes~\cite{Sachdev_2010,Kitaev_talk_2015}.
  Could these models have something to say about quantum gravity analogous to
  the SYK models?}}\BibitemShut {Stop}%
\bibitem [{\citenamefont {Sachdev}(2010)}]{Sachdev_2010}%
  \BibitemOpen
  \bibfield  {author} {\bibinfo {author} {\bibfnamefont {S.}~\bibnamefont
  {Sachdev}},\ }\bibfield  {title} {\bibinfo {title} {Holographic metals and
  the fractionalized fermi liquid},\ }\href
  {https://doi.org/10.1103/PhysRevLett.105.151602} {\bibfield  {journal}
  {\bibinfo  {journal} {Phys. Rev. Lett.}\ }\textbf {\bibinfo {volume} {105}},\
  \bibinfo {pages} {151602} (\bibinfo {year} {2010})}\BibitemShut {NoStop}%
\end{thebibliography}%

\end{document}